\documentclass[manuscript,screen,usenames,dvipsnames]{acmart}
\usepackage{booktabs} 
\usepackage[ruled]{algorithm2e} 
\usepackage{dsfont} 

\acmJournal{TOPS}
\acmVolume{9}
\acmNumber{4}
\acmArticle{39}
\acmYear{2021}
\acmMonth{3}

\usepackage{array}
\newcolumntype{x}[1]{%
>{\raggedleft\hspace{0pt}}p{#1}}%

\usepackage{calc}
\makeatletter
\newcounter{tmp@cnt}
\newcommand*\@labelpunc{}
\newcommand*\combine[1][2]{%
    \refstepcounter{enumi}
    \setcounter{tmp@cnt}{\value{enumi}}
    \addtocounter{enumi}{#1-1}
    \item[(\thetmp@cnt+\theenumi\@labelpunc)]}
\newcommand*\labeltype[2][]{\gdef\@labelpunc{#1}\renewcommand\thetmp@cnt{#2{tmp@cnt}}}
\makeatother
\usepackage{enumitem} 

\newcommand*\rot{\rotatebox{90}}
\usepackage{multirow}
\usepackage{rotating}

\setcopyright{acmcopyright} 
\usepackage{wrapfig}
\usepackage{setspace}
\usepackage{multicol}
\usepackage{supertabular}
\usepackage{wasysym}   
\usepackage{longtable}
\usepackage{tikz}
\usepackage{colortbl}
\usepackage{changepage}
\acmDOI{0000001.0000001}
\usepackage{supertabular}
\usepackage{setspace} 
\usepackage{subcaption} 
\usepackage{tikz} 
\usetikzlibrary{shapes.geometric, arrows,positioning} 

\usetikzlibrary{mindmap} 

\usepackage{caption}

\usepackage{pifont}
\usepackage{bm}

\usepackage{ amssymb } 

\usepackage{wasysym}
\usepackage{scalerel} 
\usepackage{ marvosym } 
\usepackage{longtable}
\usepackage{multirow}
\usepackage{multicol}
\usepackage{supertabular}
\usepackage{ amssymb } 

\usepackage{tikz}
\usetikzlibrary{shapes.geometric, arrows,positioning}
\usetikzlibrary{shapes,arrows,positioning,patterns,decorations.pathmorphing}
\usepackage[linguistics]{forest}
\usepackage{adjustbox}
\usepackage{array}
\usepackage{booktabs}
\usetikzlibrary{mindmap}
\usepackage{tcolorbox}
\tikzset{every node/.append style={scale=0.7}}

\usepackage{xr}

\usepackage{scalerel}

\usepackage{graphicx}

\newcommand{\f}{\newmoon}               
\newcommand{\p}{\fullmoon}              

\newcommand{\fv}{\underline{\newmoon}}               
\newcommand{\pv}{\underline{\fullmoon}}              

\newcommand{\flog}{$\hphantom{^{@}}\colourone\f^{@}$}               
\newcommand{\plog}{$\hphantom{^{@}}\colourone\p^{@}$}              

\newcommand{\fper}{$\hphantom{^{\sun}}\colourtwo\f^{\sun}$}               
\newcommand{\pper}{$\hphantom{^{\sun}}\colourtwo\p^{\sun}$}              

\usepackage{dashbox}%

\newcommand{\colourone}{\color{blue}}
\newcommand{\colourtwo}{\color{magenta}}

\newcommand{\flogv}{$\hphantom{^{@}}\colourone\underline{\f}^{@}$}               
\newcommand{\plogv}{$\hphantom{^{@}}\colourone\underline{\p}^{@}$}  

\newcommand{\fperv}{$\hphantom{^{\sun}}\colourtwo\underline{\f}^{\sun}$}               
\newcommand{\pperv}{$\hphantom{^{\sun}}\colourtwo\underline{\p}^{\sun}$}  

\newcommand{\posper}{$\hphantom{^{\sun}}\colourtwo+^{\sun}$}

\newcommand\colorboxpink[1][H]{\setlength{\fboxsep}{1pt}\colorbox{lightgray!40!pink!50!}{\phantom{#1}}}
\newcommand\colorboxblue[1][H]{\setlength{\fboxsep}{1pt}\colorbox{lightgray!40!blue!25!}{\phantom{#1}}}
\newcommand\colorboxgrey[1][H]{\setlength{\fboxsep}{1pt}\colorbox{lightgray!50!}{\phantom{#1}}}
\newcommand{\veryup}{$\color{red}{\bm{\Uparrow}}$} 
\newcommand{\up}{$\color{red}{\bf{\uparrow}}$} 
\newcommand{\verydown}{$\color{Green}{\bm{\Downarrow}}$} 
\newcommand{\down}{$\color{Green}{\bm{\downarrow}}$} 

\usepackage{placeins}

\newcommand{\X}{\color{red}{\ding{56}}\color{black}} 
\newcommand{\Y}{\color{Green}{\ding{52}}\color{black}} 
\newcommand{\n}{\color{Goldenrod}{\ding{110}}\color{black}} 
\newcommand{\splitYNo}{\color{Goldenrod}{\ding{51}/\ding{55}}\color{black}}
\newcommand{\splitYn}{\color{Goldenrod}{\ding{51}/\ding{110}}\color{black}}
\newcommand{\splitNn}{\color{Goldenrod}{\ding{110}/\ding{55}}\color{black}}

\begin{document}

\title{Costs and benefits of authentication advice} 

\author{Hazel Murray}
\orcid{0000-0002-5349-4011}
\affiliation{%
  \institution{Munster Technological University}
  \streetaddress{Bishopstown}
  \city{Co. Cork}
  \country{Ireland}}
\email{hazel.murray@mtu.ie}
\author{David Malone}
\affiliation{%
  \institution{Maynooth University}
  \city{Co. Kildare}
  \country{Ireland}
}
\email{david.malone@mu.ie}

\begin{abstract}
Authentication security advice is given with the goal of guiding users and organisations towards secure actions and practices. In this paper, we demonstrate that security advice can be ambiguous, contradictory and at times may not even have any clear benefits. We expand on current work by defining a formal approach to identifying costs of security advice and instigate a user study to identify the costs that apply to a large range of authentication advice. We also apply a simple framework for analysing the authentication related security benefits of advice. This allows us to identify costs and benefits for all classes of security advice.
\end{abstract}


%
%


\maketitle

\section{Introduction} 
Password are an essential part of our online security. However, the advice and restrictions placed on passwords have made them a source of considerable inconvenience for users~\cite{adams1999users}. Rules introduced around passwords and other authentication procedures sometimes seem unsupported by research and their security objectives can be unclear~\cite{xkcdtroub4dour}.  

In his 2003 book ``Beyond Fear'', Schneier explains that ``almost every security measure requires trade-offs. These trade-offs might be worse usability of a system, additional financial costs or a decrease of security in another place''~\cite{bruce2003beyond}. In this paper we delve deeper into the trade-offs and effects of security policies and discover whether decreases to usability are justified by security increases. 

Understanding the extent of the costs to the user or organisation and trading these off for the benefits is important if we envision a user and organisation as having a fixed amount of effort they are willing to exert for their security~\cite{beautement2009compliance}. Wasting a large amount of this \textit{compliance budget} with advice that is high cost and low impact means other advice with more effective benefits might get pushed aside. This research highlights the often-overlooked usability costs of authentication security policies. 

In this paper, we collected 270 pieces of authentication security advice that were given by security specialists, multinational companies and public bodies. We noticed stark variations between the advice that was given and 41\% of the recommendations we collected were contradictions of those given by another source. Determining whether advice is ``worthwhile'' seems to be a difficult task and one that is not well understood by those giving the advice. To learn more about the costs versus security benefits of the advice that is given, we surveyed administrators and users about the costs associated with adhering to this advice. The results of this study are tabulated and illustrate a methodology for identifying security benefits  and costs of authentication advice and highlight the difficulty of such an endeavour. This research highlights the need for organisations to follow the best practices guidelines when giving advice rather than preconceived notions of what should be secure.

The paper is structured as follows: In Section~\ref{sec:colcat} we describe our methods for collecting and categorizing authentication advice. Tab.~\ref{tab:statements} depicts the output of the coding and highlights the inconsistencies in the advice collected. In Section~\ref{sec:advice}, we select and discuss four chosen advice categories and offer insights on them with respect to current research and inconsistencies. In Section~\ref{sec:costs}, we define a method for identifying the costs associated with implementing authentication advice. Section~\ref{sec:survey_descriptions}, describes the set up of the user study which investigates how users and administrators are affected by the need to follow or implement the collected security advice. The results of this user study are presented in Tab.~\ref{tab:table_of_costs}. Section~\ref{sec:benefits} outlines our model of the benefits of authentication advice. These results are analysed and discussed in Section~\ref{sec:cost-results}. The benefits of the collected authentication advice is shown in Tab.~\ref{tab:table_of_benefits} and discussed in Section~\ref{disc_ben}. In Section~\ref{sec:costsvsbenefits} the trade-off between cost and benefits for authentication advice is discussed. We supply additional analysis and the the anonymous responses of our user study respondents in our GitHub repository~\cite{hazel_github}.  We finish with a summary of results and conclusion in Sections~\ref{sec:summ} and Section~\ref{sec:conc} respectively.

\section{Related Work}\label{sec:related work} 
Lampson \cite{lampson2009privacy} in 2009 said that ``The root cause of the problem is economics: we don’t know  the  costs  either  of  getting  security  or  of  not  having  it'' ... ``To  fix  this  we  need  to  measure  the cost   of   security,   and   especially   the time  users  spend  on  it''.

Usability with security has become a subject of interest in the last few years. It began in 1999 with the publication of Adams and Sasse's paper ``Users are not the enemy''~\cite{adams1999users}. They found that users employ workarounds for the security policies which do not align with their work procedures. Adams and Sasse emphasise that security systems must be designed with usability in mind, as otherwise mechanisms that might look secure on paper will fail in practice.

In 2005, Cranor and Garfinkel published their book ``Security and Usability: Designing Secure Systems that People Can Use'' \cite{cranor2005security}. This book covered pinnacle topics in the usable security discussion. Of particular relevance to our work is the chapter ``Evaluating Authentication Mechanisms'' by Karen Renaud.  In this chapter~\cite{renaud2005evaluating} and in the related paper~\cite{renaud2004quantifying}, Renaud quantified the quality of web authentication mechanisms. Uniquely at this time, Renaud took into account usability metrics. Renaud evaluated a selection of different authentication mechanisms with respect to accessibility, memorability, security and vulnerability.

Authentication policies place a large burden on users and organizations \cite{inglesant2010true}. In the aptly titled paper ``The true cost of unusable password policies: password use in the wild'' Inglesant and  Sasse  \cite{inglesant2010true} find that users are generally concerned with maintaining security, but that while an organization or user may want to enforce strong security, if the time and monetary constraints are too high then it might not be feasible. Beautement et al. find that bypassing security policies is a widely employed practice \cite{beautement2009compliance}. They introduce the idea of a compliance budget, which formalizes the understanding that users and organizations do not have unlimited capacity to follow new instructions and advice.

Herley~\cite{herley2009so} argues that a users' rejection of security advice is rational from an economic perspective. Herley quantifies the costs versus benefits for three specific authentication guidelines: password rules, phishing site identification advice and SSL certificate warnings. Redmiles et al. show that when participants are explicitly aware of security risks and costs they will likely make a rational (utility optimal) security choice~\cite{redmiles2018dancing}. This provided evidence that if a user perceive the policy to be beneficial to them then they  will be more likely to follow it. This suggests that we should write clear policies which emphasise a benefit that will counteract the costs users will be expected to bear.  

Braz et al. \cite{braz2007designing} propose a usability inspection method which can be used to compare the security problem to the usability criteria for online tasks such as: authenticate yourself, transfer funds or buy a concert ticket. Shay and Bertino \cite{shay2009comprehensive} and Arnell et al. \cite{arnell2012systematic} have built simulations to model the costs versus the benefits of a password policy with complexity rules, throttling and regular expiry. They hope to help organizations to determine the trade-offs between the user and organization costs and the security improvements of complexity rules. Shay and Bertino explain that security is an economic as well as a computer problem. 

Research into other areas of authentication security are also being conducted. Researchers are interested in guessing password \cite{kelley2012guess}, password reuse \cite{wang2018next} \cite{golla2018site} and alternatives to passwords \cite{bonneau2012quest} \cite{stajano2011pico}.


\section{Collection and Categorization of Password Advice}\label{sec:colcat} 

In our previous work we detail the collection of password advice \cite{murray2017evaluating} which we will briefly review as this work builds upon it. We collected 270 pieces of authentication advice from 20 different sources. We primarily used Internet searches to collect password advice but also looked at advice given by standards agencies and multinational companies. We attempted to recreate the actions an individual or organization might take when seeking to inform themselves about proper password practices.

As the advice was collected it was systematically subdivided into categories. In total, we identified 27 categories shown in Table \ref{tab:Categories}. The categories are listed in two columns, one showing categories containing advice aimed at users and the second showing advice aimed primarily towards organisations.
Also included are the number of pieces of advice under each category. We collected 179 pieces of advice aimed towards users and 91 pieces aimed towards organisations.

\begin{table}
\centering
\caption{\label{tab:Categories}Categories and the quantity of advice they contain.}
\small
{\renewcommand{\arraystretch}{1}%
\begin{tabular}{|l|c|l|c|}\hline
\rule{0pt}{1em}\textbf{Users}&\#&\textbf{Organisations}&\#\\\hline
\rule{0pt}{1em}Phrases  &39&    Expiry  &27\\
Composition &28&   Storage &16\\
Personal Information    &21&   Generated passwords &7\\
Reuse   &17&   Individual accounts &7\\ 
Personal password storage   &17&   Throttling guesses  &6\\ 
Length  &17& Keeping system safe &6\\ 
Sharing &13& Default passwords   &4\\
Keep your account safe  &10&     Administrator accounts  &4\\
Backup password options &8& Input&3\\
Password managers   &4& Shoulder surfing    &3\\
Two factor authentication   &3& Policies    &2\\
Username requirements   &2&  Transmitting passwords  &2\\
&&SNMP strings&2\\
&&Password auditing&1\\
&&Back up work&1\\\hline
\rule{0pt}{1em}\textbf{Total}&\textbf{179}&\textbf{Total}&\textbf{91}\\\hline
\end{tabular}}
\end{table}

To extract insight into the views and opinions on the advice, we created over-arching statements to describe the different views of the advice within each category. For example, the category ``Expiry'' contains 27 pieces of advice. Five encourage organizations to store previous user passwords to prevent cycling through previous passwords; 12 relate to periodic password changes; and 10 tell users and organizations to change passwords if they suspect compromise. Therefore the category ``Expiry'' contains three \underline{advice statements}: (1) Store password history to eliminate reuse, (2) Change your password regularly, (3) Change password if compromise is suspected. 
In total, we created 80 advice statements. Then, for each of these advice statements we compare them to the individual pieces of advice. For example, below are the twelve pieces of advice under the statement ``Change your password regularly'':

\begin {enumerate} 

\item Passwords should be changed periodically. 

\item Enforce a Maximum Password Age, users should change passwords regularly. 

\item Change your password regularly. 

\item Have a minimum Password Age.

\item Password expiration should be enforced on all accounts. 

\item Change your passwords regularly (every 45 to 90 days). 

\item All user-level passwords must be changed at least every six months.  

\item All system-level passwords must be changed on at least a quarterly basis. 

\setcounter{enumi}{0} 

\item *The routine changing of passwords is not recommended. 

\item *Don't change your passwords, unless you suspect they've been compromised. 

\item *Verifiers SHOULD NOT require memorized secrets to be changed arbitrarily (e.g., periodically). 

\item *Normally, there should be no reason to change your password or PIN. 

\end {enumerate}

We can see the first eight pieces of advice fit under this category because they encourage users or organizations to enforce periodic password changes. The last four pieces of advice also belong in this category because they disagree with periodic password changes. We chose to display it in this way as it highlights the contradictions in the advice circulated. This gives an insight into why organizations and users might have so much trouble following authentication recommendations.  

The first column of Tab.~\ref{tab:statements} shows each advice category and the advice statements related to each. In the second and third column of the tables we show how many pieces of advice support that advice statement, and how many pieces of advice contradict the advice statement. We use an asterisk, *, to denote a statement which is contradicted by a separate statement in the category. For example, ``Never reuse a password'' does not have advice within the statement that contradicts it. However, the third statement ``Don't reuse certain passwords'' is a contradiction of never reusing passwords.

\section{Advice discussion}\label{sec:advice}
Looking through Tab.~\ref{tab:statements}, we can see that for many categories we collected advice which was given by one source and contradicted by another. 41\% of all the advice we collected was contradicting.

In this section, we will discuss some observations that we made for four chosen categories of advice. For our insights on all the advice categories, see the supplementary material here~\cite{hazel_github}. 

\subsection{Phrases}
Advice regarding password phrases was the most commonly given advice we encountered. This implies that advice is mostly concerned with making passwords ``strong''. While this is important for some attacks, for attacks such as phishing and keylogging the strength of the password is irrelevant~\cite{zhang2016revisiting,florencio2016pushing}.

Within the category \textit{Phrases} there were no contradictions for the statements: \textit{Don't use patterns, Take initials of a phrase} and \textit{Don't use words.} The last is particularly interesting since from leaked password database we know users primarily choose word based passwords~\cite{rockyou}. Shay et al. find that the ``use of dictionary words and names are still the most common strategies for creating passwords''~\cite{shay2010encountering}. This depicts how ineffective some password advice can be and is possibly a reflection on the costs appearing to not outweigh the benefits from a users' point of view. 

The statements: \textit{Don't use published phrases} and \textit{Substitute symbols for letters} had contradictions. For \textit{don't use published phrases} the advice given was:
\begin{enumerate}
\item ``Don't use song lyrics, quotes or anything else that has been published.''
\item ``Do not choose names from popular culture.''
\item ``Choose a line of a song that other people would not associate with you.''
\end{enumerate}
The last piece of advice directly contradicts the first. This inconsistency makes it no surprise that users seem disinclined to follow security advice~\cite{inglesant2010true,adams1999users}.

The advice statement \textit{Substitute symbols for letters} is proposed by two sources but is advised against by a third. We know from Warner~\cite{subs} that passwords with simple character substitutions are weak. Yet, 2 of 3 pieces of advice recommend it. This could stem from the attitude that it is ``better than nothing''.

\subsection{Reuse}
We collected six pieces of advice telling users to \textit{never reuse passwords} and three pieces telling users to \textit{not reuse passwords for certain sites}. In addition, we found three pieces of advice encouraging users to \textit{alter and reuse their passwords} and three pieces telling users to not alter and reuse their passwords. There seems to be little agreement among the distributed advice in terms of password reuse. 

Das et al. estimate that 43--51\% of users re-use passwords across sites~\cite{das2014tangled}. They also provide algorithms that improve an attacker's ability to exploit this fact. Flor{\^e}ncio, Herley and Van Oorschot~\cite{florencio2014password} declare that, ``far from being unallowable, password re-use is a necessary and sensible tool in managing a portfolio'' of credentials. They recommend grouping passwords according to their importance and reusing passwords only within those groups. Interestingly, the advice we collected in the category \textit{Don't reuse certain passwords} gave a slightly different take on this advice. The advice mostly asked users to not use the password used for \textit{their site} anywhere else, e.g. ``Never use your Apple ID password for other online accounts''. Most organisations gave advice prioritizing their own accounts. Only one piece of advice suggested using a unique password for any important accounts~\cite{google}.

In our user study (described later), one respondent said ``I experience regular frustration and lose valuable time at work trying to login to numerous platforms with different passwords''.

\paragraph{Alter and reuse passwords}
An alternative to grouping accounts for reuse is to alter and then reuse a password. This advice was given by three sources and rejected by three sources. Alterations to password are often predictable. Using a cross-site password guessing algorithm, Das et al.~\cite{das2014tangled} were able to guess approximately 10\% of non-identical password pairs in less than 10 attempts and approximately 30\% in less than 100 attempts.

\subsection{Composition}
Composition restrictions are regularly enforced by websites but the advice given is not consistent from site to site. It is interesting to note that Herley~\cite{herley2009so} hypothesizes that different websites may deliberately have policies which are restrictive to different degrees as this can help ensure that users do not share passwords between sites. Below we will discuss one of the advice statements associated with password composition requirements: Enforce restrictions on characters.

We collected twelve pieces of advice encouraging requirements on the characters that should be used in passwords and only one piece of advice against it. The source rejecting composition rules was the NIST 2017 authentication guidelines~\cite{nist2017}. These guidelines have received praise from the authentication research community~\cite{nistcox}. This raises the question about whether organisations will begin to disseminate these new security practices or continue to enforce their stringent password restrictions.

Kelley et al. show that strict composition rules do increase the security of passwords~\cite{kelley2012guess}. However, similar security can be offered by password blocklists or mandating minimum 16-character passwords. In a related study, Komanduri et al.~\cite{komanduri2011passwords} showed that the 16-character password restriction was less annoying and less difficult for users. In addition, when composition rules are enforced, the probability of a user including a ``1'' as their number and an ``!'' as there symbol is high~\cite{ur2015added}. So, an attacker who knows the composition restriction in place could potentially tailor their attacks to suit it. Tan et al. also echoes this finding stating that requiring passwords to have multiple character classes has at most minor benefits to password strength and can even reduce effective security~\cite{tan2020practical}.

\subsection{Expiry}
We found five pieces of advice telling organisations to \textit{Store password history to eliminate reuse}, one encouraged organisations to \textit{Enforce a minimum password age} and ten were in favor of \textit{Changing passwords if compromise is suspected.}
If organisations do \textit{store their users' password history} this creates an additional security hole as the company needs to allocate resources to protect this file. Also, even though users can no longer reuse prior passwords, alterations are still possible~\cite{shay2010encountering}. In fact, Zhang, Monrose and Reiter~\cite{zhang2010security} identify that we can easily predict new passwords from old when password aging policies force updates.

The reason given for introducing a \textit{minimum password age} is to prevent users from bypassing the password expiry system by entering a new password and then changing it right back to the old one~\cite{technetmag}. However, if an attacker gains access to a users' account and changes their password the user will be unable to change it again until the required number of days have elapsed, or with an administrators' help. 

Ten pieces of advice recommended \textit{changing passwords if a compromise is suspected}. This can be inconvenient for users not affected by the compromise, and also those who are. If there is a breach at the server the users were not at fault yet still they must choose a new password.

From anecdotal evidence we know the advice \textit{change your password regularly} is widely hated by users~\cite{hatechangingpass}. This is summarized by one user in our user study saying ``I hate this! The only solution I've come up with is to increment a number in the password each time. So inconvenient and frustrating, especially when combined with other bad password advice''. 
Seven pieces of the advice we collected encouraged the use of password expiry while only four pieces of advice discouraged it. This is despite research suggesting that the security benefits are minimal~\cite{chiasson2015quantifying,zhang2010security}. This implies the inconvenience to users is worth less to organisations than the minimal security benefits. Or, do organisations want to be seen to be enforcing strong security practices, and forcing expiry is just one way of doing this?

\noindent As mentioned, more discussions of the advice collected can be found in~\cite{hazel_github}.



\begin{singlespace}
\begin{table}
\caption{Breakdown of advice into statements}\label{tab:statements}
 \begin{subtable}[t]{0.5\textwidth}
  \small
  \flushleft
 {\begin {tabular}{|l|c|c|}

\multicolumn{1}{c}{\large{Users}} &\multicolumn{1}{c}{\#x} & \multicolumn{1}{c}{\#\checkmark}\\\hline

Backup password options  && \\\hline
Email up-to-date and secure.  &0 & 3  \\
Security answers difficult to guess. &0 & 3  \\
Do not store hints.& 0 & 2  \\\hline

Composition & & \\\hline
Must include special characters  & 5 & 7 \\
Don't repeat characters. & 0&  3  \\
Enforce restrictions on characters.  & 1 & 12\\\hline

Keep your account safe  && \\\hline
Check web pages for SSL&0 &1   \\
Manually type URLs. &0 & 1  \\
Don't open emails from strangers.& 0 & 1  \\
Keep software updated. &0 & 2  \\
Keep anti virus updated.  &0 &2   \\
Log out of public computers. &0 & 2  \\
Password protect your phone.&0 & 1  \\\hline

Length  && \\\hline
Minimum password length.&0 &13   \\
Enforce maximum length (\textless40).&1 &3   \\\hline

Password Managers  && \\\hline
Use a password manager.  &1 &2   \\
Create long random password. &0 & 1  \\\hline

Personal Information  & & \\\hline
Don't include personal information.&1 & 5 \\
Must not match account details.&0 & 8  \\
Do not include names. & 1 & 6  \\\hline

Personal password storage  && \\\hline
Don't leave in plain sight.  &0 &4   \\
Don't store in a computer file. &1 & 2  \\
Write down safely.& 1 & 6  \\
Don't choose ``remember me''. &0 & 3  \\\hline

Phrases && \\\hline
Don't use patterns.  &0 & 6  \\
Take initials of a phrase. &0 & 4  \\
Don't use published phrases. & 1 & 2  \\
Substitute symbols for letters. &1 & 2  \\
Blocklist common passwords &0&2 \\
Don't use words.  &0 & 16  \\
Insert random numbers and symbols &1 & 4\\\hline

Reuse & & \\\hline 
Never reuse a password. &*5& 6  \\
Alter and reuse passwords& 3 &  3  \\
Don't reuse certain passwords.  & 0 & 5 \\\hline

Sharing && \\\hline
Never share your password.&0 &9   \\
Don't send passwords by email.&0 &3   \\
Don't give passwords over phone.&0 &1   \\\hline

Two factor authentication && \\\hline
Use for remote	 accounts.  &0 &1   \\
Use multi-factor authentication.  &0 &1   \\
2 factor authentication using phone.  &0 &1   \\\hline

Username  & & \\\hline
Enforce composition restrictions.&0 &1   \\
Don't reuse username. &0 & 1  \\\hline

\end {tabular}}
\caption{User advice statements}
\centering{\shortstack{\#x = number sources contradicting.\\\#\checkmark= number sources advising.}}
\end{subtable}%
 \begin{subtable}[t]{0.5\linewidth}
 \small
 \flushright
 {\begin {tabular}{|l|c|c|}

\multicolumn{1}{c}{\large{Organisation}} &\multicolumn{1}{c}{\#x} & \multicolumn{1}{c}{\#\checkmark}\\\hline

Administrator accounts && \\\hline
Not for everyday use.&0 &1   \\
Must have it's own password.&0&2   \\
Should have extra protection.&0&1   \\\hline

Backup work && \\\hline
Make digital \&\ physical back-ups.  &0 &1   \\\hline

Default passwords && \\\hline
Change all default passwords.&0 &4   \\\hline

Expiry  && \\\hline
Store history to eliminate reuse.  &0 &5   \\
Change your password regularly.&4 &8   \\
Change if suspect compromise.&0 &10   \\\hline

Generated passwords && \\\hline
Use random bit generator.  &*2 &2   \\
Must aid memory retention.  &0 &2   \\
Must be issued immediately.  &0 &1   \\
Only valid for first login.  &0 &1   \\
Distribute in a sealed envelope.&0 &1   \\\hline

Individual accounts  && \\\hline
One account per user.&0 &4   \\
Each account password protected.&0&3   \\\hline

Input && \\\hline
Don't performed truncation.&0 &1   \\
Accept all characters.&1 &1   \\\hline

Keeping system safe && \\\hline
Implement Defense in Depth.  &0 &2   \\
Implement Technical Defenses.  &0 &1   \\
Apply boot protection.&0 &1   \\
Monitor and analyse intrusions.&0 &1   \\
Regularly apply security patches.  &0 &1   \\\hline

Network: SNMP community strings && \\\hline
\shortstack{Don't define as standard defaults.}&0 &1   \\
Different to login password.&0 &1   \\\hline

Password auditing  && \\\hline
Attempt to crack passwords.&0 &1   \\\hline

Policies && \\\hline
Establish clear policies.&0 &2   \\\hline

Shoulder surfing  && \\\hline
Offer to display password.&0 &1   \\
Enter your password discretely.&0& 2   \\\hline

Storage && \\\hline
Encrypt passwords.&*4& 7   \\
Restrict access to password files.&0 &2  \\
Encrypt password files.  &0 &1   \\
Store password hashes.&0 &4   \\
Don't hardcode passwords.&0 &1  \\
Contracts state how pwds protected. &0 &1\\\hline

Throttling && \\\hline
Throttle password guesses.&0 &6   \\\hline

Transmitting passwords && \\\hline
Don't transmit in cleartext.&0 &1   \\
Request over a protected channel.&0&1   \\\hline

\end {tabular}}
\caption{Verifier advice statements}
\end{subtable}
\end{table}
\end{singlespace}


\section{Costs Model}\label{sec:costs} 

In this section, we describe our methodology for the creation of our costs model. We began with a brainstorming exercise. At the heart of it was a conscious consideration for the usability costs: costs which are often-overlooked when security policies are implemented~\cite{cranor2005security}. We viewed costs as any burdens on the user who must follow the advice, or any burdens on the organisation who must either follow or enforce the advice.

\subsection{Cost categories}
The finalised set of costs were reached through an iterative process which involved assigning costs to each of our 79 advice statements and adapting the categories until we reached a set of cost categories that made it possible to describe the range of advice impacts. This finalized list of costs, as shown in Tab.~\ref{tab:list cost cats}, were then presented to user and administrator participants in our study for feedback.  
  


\subsubsection{User versus organisation costs}\label{subsec:u v c} 
Analyzing the cost categories, we wanted some way to distinguish between costs borne by the user and costs borne by the organisation. We believe it would be interesting to know who bears most of the costs: the user or the organisation. In addition, what might be a small cost for an organisation could be a large cost for a user. We therefore separated cost categories to be user and organisation specific e.g. \textit{user computing power}  separate from \textit{organisation computing power}.

\subsubsection{Minor costs}\label{subsec:partial costs} 
In our analysis we acknowledged the need to distinguish the extent to which a cost occurs. For example, both ``Enforce restrictions on characters in passwords'' and ``Make digital and physical back ups'' require organisation time. But little organisation time is needed to enforce a composition restriction and a lot of effort is needed to digitally and physically back up work. It is not within the scope of this model to have a full grading of the costs, but we do acknowledge a difference between small or partially felt costs and more substantial costs. In our tables a minor cost is identified as \fullmoon, and a more substantial cost is identified as \newmoon.

\subsubsection{Periodic costs}\label{subsec:periodic costs} 

We also notice that advice such as ``Make digital and physical back ups'' and ``Keep email up-to-date and secure'' require continued action. This significantly increases the costs. We acknowledge three types of costs: once off costs usually relating to setup or account creation, costs which occur at every login and periodic costs which occur repeatedly over different time frames. To identify these costs, we use a super-scripted symbol. Costs at login are denoted by an at symbol: $\colourone\newmoon^{@}$ and periodic costs are denoted by a sun: $\colourtwo\newmoon^{\sun}$. 

\subsubsection{Positive costs}\label{subsec:positive costs} 

Finally, some pieces of advice \emph{reduced} the burden on the organisation or user. These pieces of advice we acknowledge as `positive costs' and as such they are represented with the positive symbol: +. 

\begin{table} 

\caption{\label{tab:list cost cats} Finalized cost categories.}

\centering 

\begin{tabular}{l}\hline 

\textbf{Organisation Costs}\\\hline 

Increased help desk/user support time\\

User education required\\

Organisation needs extra resources\\

Takes organisation time to implement\\

Increases the organisation's computing power needed\\\hline

\textbf{User Costs}\\\hline 

Makes it more difficult to create a password\\

Makes it less easy to remember\\

Requires extra resources\\

Requires the creation of a new password\\

Increases the computing power needed\\

Requires other extra time or effort\\\hline
\end{tabular} 
\end{table}

 \subsection{User and administrator surveys}\label{sec:survey_descriptions}
Once satisfied that we had a provisional set of cost categories for users and for organisations we began our user study. This involved a set of surveys aimed at administrators and a set aimed at users. There were 10 surveys in total, five aimed at administrators and five aimed towards end users.  These are detailed below. 

We created a series of surveys which asked users and administrators to identify which categories of costs apply to each piece of advice. Participants were randomly redirected to one of the 5 surveys relevant to them, each survey contained questions about a subset of the advice. This allowed us to ask about the large number of advice statements we had collected without overburdening participants. We did include some overlap in the end-user surveys as every user was asked about: 
\begin{itemize}
    \item Composition - ``include specific character types in your password.''
    \item Expiry - ``change your password regularly''
    \item Reuse - ``never reuse a password''
    \item Sharing - ``never share your password''
    \item Paste ``you should not be able to paste your password when logging in''
    \item Password manager - ``you should use a password manager to store your passwords''
    \item Two-factor authentication - ``you should use 2-factor authentication when logging into accounts (e.g. a code from an app or a special device)''
\end{itemize}
Note that the advice ``Do not allow users to paste passwords'' was not collected as part of our research. Instead we chose to add it into our survey as a ``canary''. This advice has been discussed online as a piece of bad advice which unfortunately became common practice, but has no discernible security ramifications \cite{troypaste}. This advice can act as a test case for our methods. 

There was no overlap in the administrators' surveys as they were asked a larger variety of questions. For example, administrators were quizzed on costs associated with hashing and salting passwords and monitoring and analysing intrusions which users were not asked about.

Participants were chosen using a snowball sampling technique. For a full description of the survey methodology and ethical considerations, please see~\cite{murray2021improving}. The study was approved by our university Ethics Review Board.

\begin{figure}
    \centering
    \includegraphics[width=0.6\textwidth]{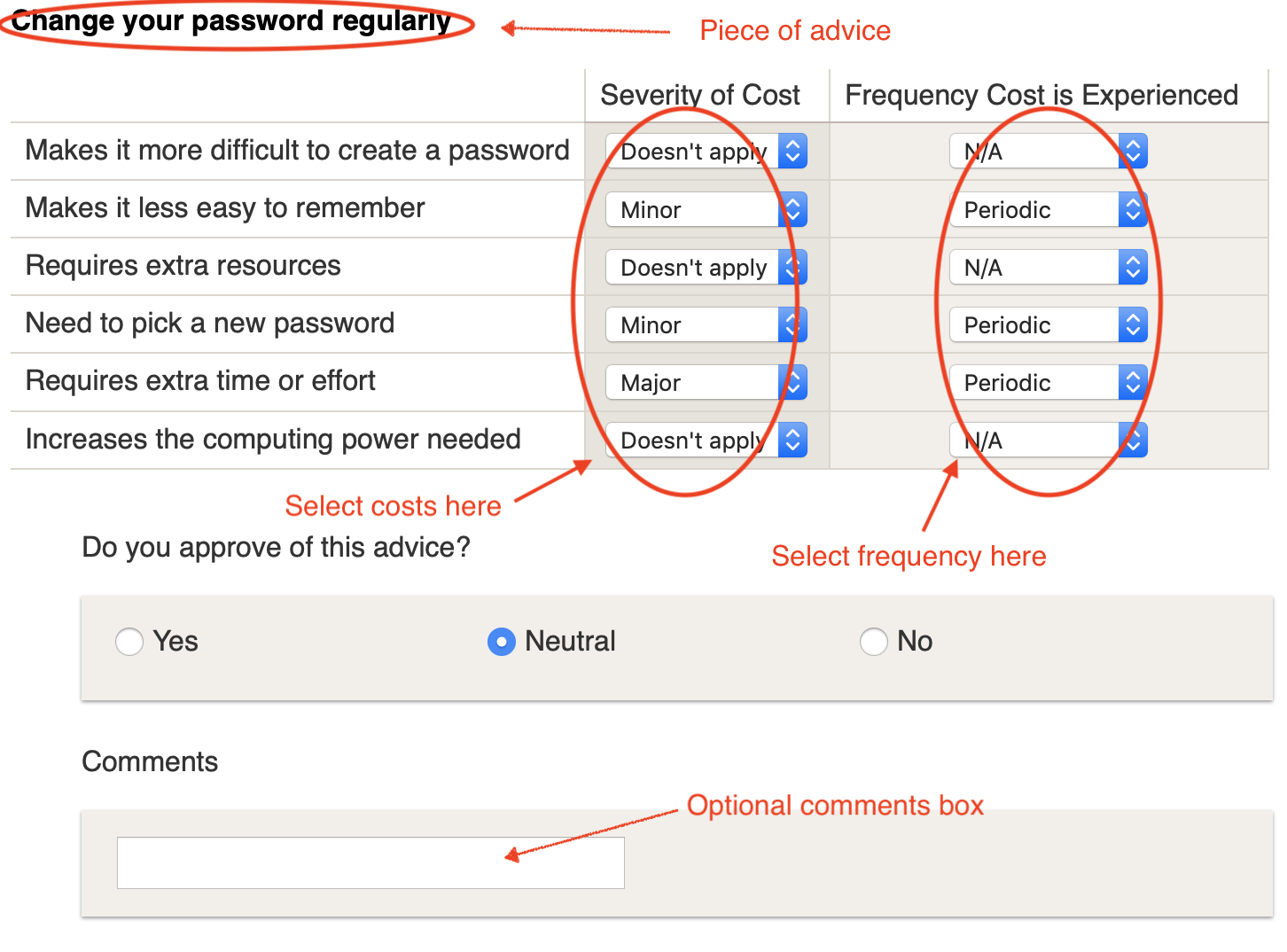}
    \caption{Infographic with instructions for users for the survey}
    \label{fig:instr_user_survey2}
\end{figure}

In the survey, participants were asked to indicate the severity and frequency with which they experience costs and inconveniences as a result of authentication advice. We also asked for comments about whether they approved of the given piece of advice. Fig.~\ref{fig:instr_user_survey2} shows a graphic providing users with an explanation of how to complete a survey question. Users were also given the following example to help them understand the survey: 
\begin{quote}
For example: for the advice \textit{``Change your password regularly''} you might consider the following costs:
\begin{itemize}
    \item Makes it less easy to remember (periodically - every time I need to change it)
    \item Need to pick a new password (periodically - every time I need to change it)
    \item Takes extra time (periodically - sometimes I can't start work until I change the password)
\end{itemize}
You could complete the question for this advice as shown in Fig.~\ref{fig:instr_user_survey2}.

Note these costs are individual and may differ for you and you are not required to offer explanations.
\end{quote}
Administrators were given a similar description. Full versions of the questions and answers for both the user survey  and administrator survey are available on github~\cite{hazel_github}.

\subsection{Representation of survey responses}\label{sec: costs table discussion} 
When collecting the survey results, we use the following rules for identifying what costs the participants agreed exist for each piece of advice (i.e. did participants believe a cost major, minor, positive or non-applicable):  
\begin{itemize}
    \item If one answer option was chosen by more participants than any other then it is the majority answer.
    \item If two answers are selected with equal frequency we choose the one that represents the largest cost but note with an underline that there was variability in the responses, e.g. $\underline\newmoon$. This means we are choosing to overestimate rather than underestimate costs.
    \item If `Doesn't apply' is outweighed or equaled by minor and major combined then we assign it as minor.\footnote{Note that ``Doesn't apply'' was the default response for the survey questions.} We again indicate that there is variability. For example, if 40\% of the users said non-applicable but 30\% said it was a major cost and 30\% said it was a minor cost, then we would mark it as a minor cost as the sum of minor and major is greater than `Doesn't apply'.
\end{itemize}

Users and administrators were also asked about their approval of the advice statements. The results of this are represented in the table for both users and administrators. A \Y \hphantom{v} represents the majority of respondents' approval of the advice. \X \hphantom{v}indicates that the majority disapproved and \n \hphantom{v}represents the majority of respondents indicating that they were neutral about the advice. Occasionally, respondents were split between different approval ratings. This is also indicated in the table. For example, if users were split between approving of the advice and feeling neutral about the advice, we represent this as \splitYn.

In Tab.~\ref{tab:table_of_costs} (pages \pageref{tab:table_of_costs}--\pageref{coststableendpageref}) we present the costs that users and/or administrators associated with each piece of advice. Note, we did not include every piece of advice in the surveys in an attempt to reduce the size of the survey. In these cases, we highlight the corresponding row in the table in blue (See the organisation side of the row ``Enforce maximum length''). At times we indicate obvious costs if they exist. For example for the ``Enforce maximum length'' advice, the administrative costs could be extrapolated from the ``Minimum password length'' costs.


\thispagestyle{empty}
\normalsize
\renewcommand{\tabcolsep}{2pt}{
\setlength\LTleft{-2cm}
\begin{longtable}{lrr|c|c|c|c|c|c||c|c|c|c|c|c|c||}
\caption{\label{tab:table_of_costs}Costs of implementing password advice}
\\\cline{4-16}
\multicolumn{1}{c}{\hphantom{This is: NOTA drillwene edspace}} &\multicolumn{2}{c}{}& \multicolumn{6}{|c|}{Organisation} & \multicolumn{6}{c}{User costs} & \multicolumn{1}{c|}{}\\\cline{4-16}
\textit{Advice to Users} & \rot{\#\ sources contradicting} & \rot{\#\ sources advising} & \rot{\shortstack[l]{Increased help desk/user\\support time needed}} & \rot{\shortstack[l]{User education required}} & \rot{\shortstack[l]{Requires extra resources}} &
\rot{\shortstack[l]{Takes time to implement}} &\rot{\shortstack[l]{Increased computing\\power needed}}&
\rot{\shortstack[l]{\textit{Administrator approval}}}
& \rot{\shortstack[l]{Makes it more difficult\\to create a password}} & \rot{\shortstack[l]{Makes it less easy\\to remember}}& \rot{\shortstack[l]{Requires extra resources}}  & \rot{\shortstack[l]{Requires the creation\\of a new password}}
& \rot{\shortstack[l]{Increases the computing\\power needed}} & \rot{\shortstack[l]{Requires other extra time\hphantom{v}\\or effort}} &
\rot{\shortstack[l]{\textit{User approval}}}\\\hline
\endfirsthead
\cline{4-16}
\multicolumn{1}{c}{\hphantom{This is: NOTA drillwene edspace}} &\multicolumn{2}{c}{}& \multicolumn{6}{|c|}{Organisation} & \multicolumn{6}{c}{User costs} & \multicolumn{1}{c|}{}\\\cline{4-16}
 &\rot{\#\ sources advising} & \rot{\#\ sources contradicting}& \rot{\shortstack[l]{Increased help desk/user\\support time needed}} & \rot{\shortstack[l]{User education required}} & \rot{\shortstack[l]{Requires extra resources}} &
\rot{\shortstack[l]{Takes time to implement}} &\rot{\shortstack[l]{Increased computing\\power needed}}&
\rot{\shortstack[l]{\textit{Administrator approval}}}
& \rot{\shortstack[l]{Makes it more difficult\\to create a password}} & \rot{\shortstack[l]{Makes it less easy\\to remember}}& \rot{\shortstack[l]{Requires extra resources}}  & \rot{\shortstack[l]{Requires the creation\\of a new password}}
& \rot{\shortstack[l]{Increases the computing\\power needed}} & \rot{\shortstack[l]{Requires other extra time\hphantom{v}\\or effort}} &
\rot{\shortstack[l]{\textit{User approval}}}\\\hline
\endhead

\multicolumn{1}{l}{\textbf{Backup password options}}&   & \multicolumn{1}{c}{}&\multicolumn{1}{c}{\hphantom{\fper}}&\multicolumn{1}{c}{\hphantom{\fper}}&\multicolumn{1}{c}{\hphantom{\fper}}&\multicolumn{1}{c}{\hphantom{\fper}}&\multicolumn{1}{c}{\hphantom{\fper}}&\multicolumn{1}{c}{\hphantom{\fper}}&\multicolumn{1}{c}{\hphantom{\fper}}&\multicolumn{1}{c}{\hphantom{\fper}}&\multicolumn{1}{c}{\hphantom{\fper}}&\multicolumn{1}{c}{\hphantom{\fper}}&\multicolumn{1}{c}{\hphantom{\fper}}&\multicolumn{1}{c}{\hphantom{\fper}} &\multicolumn{1}{c|}{\hphantom{\fper}}\\\hline
Email up-to-date and secure  &0 & 3       & &	& & &  & \Y &    &       &       &     &  & \pperv& \Y  \\
Security answers difficult to guess  &0 & 3& \pper	&\fper	& 	& \pv	& & \Y   & \pv & \pv & & & & \pv & \splitYn \\
Do not store hints  &    0 & 2           &  \pper & \fperv &    &   &  &   \Y & & \flog      & &     & & \pv & \X\\\hline

\multicolumn{1}{l}{\textbf{Composition}}&   & \multicolumn{1}{c}{}& \multicolumn{12}{c}{} &\multicolumn{1}{c|}{} \\\hline

Must include special characters     & 5 &  7  &\cellcolor{cyan!50} &\cellcolor{cyan!50} &\cellcolor{cyan!50} &\cellcolor{cyan!50} &\cellcolor{cyan!50} & \cellcolor{cyan!50}&\cellcolor{cyan!50} &\cellcolor{cyan!50} &\cellcolor{cyan!50} &\cellcolor{cyan!50} &\cellcolor{cyan!50} &\cellcolor{cyan!50} &\cellcolor{cyan!50} \\
Don't repeat characters             & 0 &  3 &\cellcolor{cyan!50} &\cellcolor{cyan!50} &\cellcolor{cyan!50} &\cellcolor{cyan!50} &\cellcolor{cyan!50} & \cellcolor{cyan!50}&\cellcolor{cyan!50} &\cellcolor{cyan!50} &\cellcolor{cyan!50} &\cellcolor{cyan!50} &\cellcolor{cyan!50} &\cellcolor{cyan!50} &\cellcolor{cyan!50} \\ 
Enforce restrictions on characters   & 1 & 12 & \p & \pper & & \p & & \splitYn & \p & \plog &  &     &  & \pv & \Y \\\hline

\multicolumn{1}{l}{\textbf{Keep your accounts safe}}&   & \multicolumn{1}{c}{}& \multicolumn{12}{c}{} &\multicolumn{1}{c|}{} \\\hline
Check web pages for TLS     &0 &1  & \pperv & \fperv & & & & \Y & & & & & & \plog & \Y\\
Manually type URLs  &0 & 1   & \pper	&\pper	& & & & \X   & &  & \p& & & \fper & \n \\
Don't open emails from strangers  &0 & 1   &  \fperv & \fper & \pperv & \pper & & \Y & & & & & & \pv & \Y\\
Keep software updated  &0 & 2  &\pper &	\pper & \pperv & \fper &  & \Y & & & & & \pv & \pperv & \Y\\
Keep anti virus updated   &0 & 2 & 	&\pper	& & & & \Y   & &  & \fper & & \pper & \pperv & \Y \\
Log out of public computers   &0 & 2 & & \fper & & & & \Y & & & & & & \plogv & \Y\\
Password protect your phone   &0 & 1   & \pperv & \pper & & \p & & \Y & & \p & \p & \p & & \plog & \Y\\\hline

\multicolumn{1}{l}{\textbf{Length}}&   & \multicolumn{1}{c}{}& \multicolumn{12}{c}{} &\multicolumn{1}{c|}{} \\\hline
Minimum password length &0 &13   & \pv & \p & & \p & & \Y & \fv & \flog & & \pv & & & \Y \\
Enforce maximum length (\textless40) &1 &3 & \cellcolor{cyan!50}{\pv} &\cellcolor{cyan!50}{ \p }&\cellcolor{cyan!50} &\cellcolor{cyan!50}{\p} &\cellcolor{cyan!50} & \cellcolor{cyan!50}& \p & & & & & & \X\\\hline

\multicolumn{1}{l}{\textbf{Password managers}}&   & \multicolumn{1}{c}{}& \multicolumn{12}{c}{} &\multicolumn{1}{c|}{} \\\hline
Use a password manager  &1 &2    &\pper & \pper & \pv & \p &  & \Y & & + & \p & & & & \Y\\
Create long random password   &0 &1  & \pv & \p & & \pv & & \Y & & \f & & & & & \n \\\hline

\multicolumn{1}{l}{\textbf{Personal information}}&   & \multicolumn{1}{c}{}& \multicolumn{12}{c}{} &\multicolumn{1}{c|}{} \\\hline
Don't include personal information &1 & 5 & \cellcolor{cyan!50}&\cellcolor{cyan!50}{ \pper }&\cellcolor{cyan!50}&\cellcolor{cyan!50}&\cellcolor{cyan!50}& \cellcolor{cyan!50}& & \plog & & & & & \Y\\
Must not match account details  &0 & 8 &  & \pv & & \p & & \Y & & \plog & & & & & \Y \\
Do not include names      &1 & 6  &\cellcolor{cyan!50} &\cellcolor{cyan!50} &\cellcolor{cyan!50} &\cellcolor{cyan!50} &\cellcolor{cyan!50} & \cellcolor{cyan!50}&\cellcolor{cyan!50} &\cellcolor{cyan!50} &\cellcolor{cyan!50} &\cellcolor{cyan!50} &\cellcolor{cyan!50} &\cellcolor{cyan!50} &\cellcolor{cyan!50} \\\hline

\multicolumn{1}{l}{\textbf{Personal password storage}}&   & \multicolumn{1}{c}{}& \multicolumn{12}{c}{} &\multicolumn{1}{c|}{} \\\hline
Don't leave in plain sight &0 &4 & & \pper & & & & \Y & & \pv & & & & & \Y\\
Don't store in a computer file &1 &2  & \pperv & \pper &  & & & \Y & & \plogv & & & & \pv & \Y \\
Write down safely &1 &6  &\cellcolor{cyan!50} & \cellcolor{cyan!50}{\pper} &\cellcolor{cyan!50} &\cellcolor{cyan!50} &\cellcolor{cyan!50} & \cellcolor{cyan!50}& & + & \p & & & \p & \Y\\
Don't choose ``remember me'' &0 &3   & & \p & & & & \splitYn &  & \flogv & & & & \flog & \n\\\hline

\multicolumn{1}{l}{\textbf{Phrases}}&   & \multicolumn{1}{c}{}& \multicolumn{12}{c}{} &\multicolumn{1}{c|}{} \\\hline
Don't use patterns  &0 & 6 &\cellcolor{cyan!50} & \cellcolor{cyan!50}{\pper} & \cellcolor{cyan!50}& \cellcolor{cyan!50}{\p} &\cellcolor{cyan!50} & \cellcolor{cyan!50}& \p & \plog & & & & & \Y\\
Blocklist common password   &0 & 2  & & \pper & & \p & & \Y & \p & \p & & \pv & & \p & \Y \\
Take initials of a phrase          &0 & 4&\cellcolor{cyan!50} &\cellcolor{cyan!50} &\cellcolor{cyan!50} &\cellcolor{cyan!50} &\cellcolor{cyan!50} & \cellcolor{cyan!50}&\cellcolor{cyan!50} &\cellcolor{cyan!50} &\cellcolor{cyan!50} &\cellcolor{cyan!50} &\cellcolor{cyan!50} &\cellcolor{cyan!50} &\cellcolor{cyan!50} \\
Don't use published phrases   &1 & 2  & \cellcolor{cyan!50}& \cellcolor{cyan!50}{\pper} &\cellcolor{cyan!50} & \cellcolor{cyan!50}{\p} &\cellcolor{cyan!50} & \cellcolor{cyan!50}& \p & \flog &  & \pv & & \p & \Y\\
Substitute symbols for the letters  &1 & 2 &\cellcolor{cyan!50} & \cellcolor{cyan!50}{\pper} &\cellcolor{cyan!50} & \cellcolor{cyan!50}& \cellcolor{cyan!50}&\cellcolor{cyan!50} & \p & \plogv & & & & & \n\\
Don't use dictionary words &0 & 16  & \pper & \pper & & \p & & \Y & \fv & \flog & \p & & & \pv & \n\\
Insert random numbers and symbols  &1 & 4 &\cellcolor{cyan!50} &\cellcolor{cyan!50} &\cellcolor{cyan!50} &\cellcolor{cyan!50} &\cellcolor{cyan!50} & \cellcolor{cyan!50}&\cellcolor{cyan!50} &\cellcolor{cyan!50} &\cellcolor{cyan!50} &\cellcolor{cyan!50} &\cellcolor{cyan!50} &\cellcolor{cyan!50} &\cellcolor{cyan!50} \\\hline

\multicolumn{1}{l}{\textbf{Reuse}}&   & \multicolumn{1}{c}{}&\multicolumn{1}{c}{\hphantom{\fper}}&\multicolumn{1}{c}{\hphantom{\fper}}&\multicolumn{1}{c}{\hphantom{\fper}}&\multicolumn{1}{c}{\hphantom{\fper}}&\multicolumn{1}{c}{\hphantom{\fper}}&\multicolumn{1}{c}{\hphantom{\fper}}&\multicolumn{1}{c}{\hphantom{\fper}}&\multicolumn{1}{c}{\hphantom{\fper}}&\multicolumn{1}{c}{\hphantom{\fper}}&\multicolumn{1}{c}{\hphantom{\fper}}&\multicolumn{1}{c}{\hphantom{\fper}}&\multicolumn{1}{c}{\hphantom{\fper}} &\multicolumn{1}{c|}{\hphantom{\fper}}\\\hline
Never reuse a password  &*5 & 6 & & \fper & & & & \Y & \fperv & \flog & & \fperv &  & & \Y \\
Alter and reuse passwords& 3 & 3  & \pper & \pper & \pv & \pv & & \Y & \pper & \posper & & & & & \X \\
Don't reuse certain passwords & 0 & 5 & & \pper & & & & \Y & \pperv & \fv & & \pperv & & & \Y\\\hline

\multicolumn{1}{l}{\textbf{Sharing}}&   & \multicolumn{1}{c}{}&\multicolumn{1}{c}{\hphantom{\fper}}&\multicolumn{1}{c}{\hphantom{\fper}}&\multicolumn{1}{c}{\hphantom{\fper}}&\multicolumn{1}{c}{\hphantom{\fper}}&\multicolumn{1}{c}{\hphantom{\fper}}&\multicolumn{1}{c}{\hphantom{\fper}}&\multicolumn{1}{c}{\hphantom{\fper}}&\multicolumn{1}{c}{\hphantom{\fper}}&\multicolumn{1}{c}{\hphantom{\fper}}&\multicolumn{1}{c}{\hphantom{\fper}}&\multicolumn{1}{c}{\hphantom{\fper}}&\multicolumn{1}{c}{\hphantom{\fper}} &\multicolumn{1}{c|}{\hphantom{\fper}}\\\hline
Never share a password   &0 &9  & \pv & \pper & &  & & \Y & & & & & & & \Y \\
Don't send passwords by email  &0 &3  & & \pper & & & & \Y & & & & & & & \Y \\
Don't give passwords over phone  &0 &1 & \pv & \pper & & & & \Y & & & \p & & & & \Y\\\hline

\multicolumn{1}{l}{\textbf{Two-factor authentication (2FA)}}&   & \multicolumn{1}{c}{}& \multicolumn{12}{c}{} &\multicolumn{1}{c|}{} \\\hline
Use 2FA using app or special device  &0 &1  & \fperv & \fper & & \f & & \Y & &  & \flogv & & & \plogv & \Y \\
Use 2FA on phone &0 &1 &  \fv & \fper & \fv & \f  & \p & \splitYn & & & \flogv & &  & \pv & \Y\\
Use 2FA for remote accounts &0 &1  &\pper & \pper & \pper & \f & \pv & \Y & & & \plogv & & & \plog & \splitYNo\\\hline

\multicolumn{1}{l}{\textbf{Username}}&   & \multicolumn{1}{c}{}& \multicolumn{12}{c}{} &\multicolumn{1}{c|}{} \\\hline
Enforce restrictions on characters &0 &1  & \pper & \pper & & \pv &  & \Y & & & & & & & \X\\
Don't reuse username &0 &1 & & \pv & & & & \X & & \f &\pv & & & \fper & \splitNn \\\hline

\multicolumn{16}{l}{}\\
\multicolumn{16}{l}{\textit{Advice to organisations}}
\\\hline

\multicolumn{1}{l}{\textbf{Administrator Accounts}}&   & \multicolumn{1}{c}{}& \multicolumn{12}{c}{} &\multicolumn{1}{c|}{} \\\hline
Not for everyday use &0 &1 & \pv & \pper &  & \pv & & \Y  &\cellcolor{cyan!50} &\cellcolor{cyan!50} &\cellcolor{cyan!50} &\cellcolor{cyan!50} &\cellcolor{cyan!50} &\cellcolor{cyan!50} &\cellcolor{cyan!50} \\
Must have it's own password &0&2 & \pperv & \pv & & \p & & \Y  &\cellcolor{cyan!50} &\cellcolor{cyan!50} &\cellcolor{cyan!50} &\cellcolor{cyan!50} &\cellcolor{cyan!50} &\cellcolor{cyan!50} & \cellcolor{cyan!50}\\
Should have extra protection  &0&2 & & & & \p & & \Y  &\cellcolor{cyan!50} &\cellcolor{cyan!50} &\cellcolor{cyan!50} &\cellcolor{cyan!50} &\cellcolor{cyan!50} &\cellcolor{cyan!50} & \cellcolor{cyan!50}\\\hline

\multicolumn{1}{l}{\textbf{Backup work}} &   & \multicolumn{1}{c}{}& \multicolumn{12}{c}{} &\multicolumn{1}{c|}{} \\\hline
Make digital \& physical back-ups &0 &1 &  & \pv & \f & \p & \p & \Y & & & \fperv & &  & & \Y\\\hline

\multicolumn{1}{l}{\textbf{Default passwords}}&   & \multicolumn{1}{c}{}& \multicolumn{12}{c}{} &\multicolumn{1}{c|}{} \\\hline
Change all default passwords &0 &4 & \pper & \fv & \pv & \p & & \Y & & & & \f & & & \Y\\\hline

\multicolumn{1}{l}{\textbf{Expiry}}&   & \multicolumn{1}{c}{}& \multicolumn{12}{c}{} &\multicolumn{1}{c|}{} \\\hline
Store history to eliminate reuse &0 &5  & & \p & \p & \p & & \Y & \pper & \fper & & \fper & & & \Y \\
Change your password regularly  &4 &8 & \fper & \fperv & & \f & & \X & \fperv & \fper & & \fper & & \fper & \X\\
Change if suspect compromise &0 &10 & \pper & \pper & \pv & \pv & & \Y & & \pv & & \pv & & & \Y\\\hline

\multicolumn{1}{l}{\textbf{Generate passwords}}&   & \multicolumn{1}{c}{}& \multicolumn{12}{c}{} &\multicolumn{1}{c|}{} \\\hline
Use random bit generator &*2 &2  & \pv & & \pv & \p & \p & \Y & & \flogv & & & & & \Y\\
Must aid memory retention  &0 &2 & & & & & & \splitYn & \f & \plogv & \p & \pv & & & \n \\
Must be issued immediately &0 &1 & \pv & & & & & \Y & & \flog &  & &  & & \Y\\
Distribute in a sealed envelope  &0 &1  & & & & \pv & & \Y & & \plog & & \pv & & \p & \Y\\
Only valid for first login   &0 &1  & \pv & \p & & \p & & \Y & & & & \pv & & & \Y\\\hline

\multicolumn{1}{l}{\textbf{Individual accounts}}&   & \multicolumn{1}{c}{}& \multicolumn{12}{c}{} &\multicolumn{1}{c|}{} \\\hline
One account per user  &0 &4  & \p & \p & & \p & \p & \Y & & & & & & & \Y\\
Each account password protected &0&3 & \pv & & & & & \Y & & \pv & &\pv & & & \Y \\\hline

\multicolumn{1}{l}{\textbf{Input}}&   & \multicolumn{1}{c}{}& \multicolumn{12}{c}{} &\multicolumn{1}{c|}{} \\\hline
Don't perform truncation &0 &1  & & & \pv & \p & & \splitYn  &\cellcolor{cyan!50} &\cellcolor{cyan!50} &\cellcolor{cyan!50} &\cellcolor{cyan!50} &\cellcolor{cyan!50} &\cellcolor{cyan!50} &\cellcolor{cyan!50} \\
Accept all ASCII characters  &1 &1 & & \pv & & \pv & & \Y  &\cellcolor{cyan!50}{+} &\cellcolor{cyan!50} &\cellcolor{cyan!50} &\cellcolor{cyan!50} &\cellcolor{cyan!50} &\cellcolor{cyan!50} & \cellcolor{cyan!50}\\\hline

\multicolumn{1}{l}{\textbf{Keep system safe}}&   & \multicolumn{1}{c}{}& \multicolumn{12}{c}{} &\multicolumn{1}{c|}{} \\\hline
Implement Defense in Depth   &0 &2 & \fperv & \fperv & \fper & \fper & \p & \Y &\cellcolor{cyan!50} &\cellcolor{cyan!50} &\cellcolor{cyan!50} &\cellcolor{cyan!50} &\cellcolor{cyan!50} &\cellcolor{cyan!50} &\cellcolor{cyan!50} \\
Implement Technical Defenses  &0 &1 & & \pv & \p & \pv & \p & \Y  &\cellcolor{cyan!50} &\cellcolor{cyan!50} &\cellcolor{cyan!50} &\cellcolor{cyan!50} &\cellcolor{cyan!50} &\cellcolor{cyan!50} &\cellcolor{cyan!50} \\
Apply access control systems  &0 &1 & \pperv & \pperv & \pper & \pperv & & \Y & & & \p & & & \p & \Y\\
Monitor and analyze intrusions    &0 &1 & \pperv & \pperv & \fper & \f & \pper & \Y  &\cellcolor{cyan!50} &\cellcolor{cyan!50} &\cellcolor{cyan!50} &\cellcolor{cyan!50} &\cellcolor{cyan!50} &\cellcolor{cyan!50} &\cellcolor{cyan!50} \\
Regularly apply security patches    &0 &1 & \pv & \pper & \fper & \fper & \pperv & \Y &\cellcolor{cyan!50} &\cellcolor{cyan!50} &\cellcolor{cyan!50} &\cellcolor{cyan!50} &\cellcolor{cyan!50} &\cellcolor{cyan!50} &\cellcolor{cyan!50} \\\hline

\multicolumn{1}{l}{\textbf{Network: SNMP community strings}}&   & \multicolumn{1}{c}{} &\multicolumn{1}{c}{\hphantom{\fper}}&\multicolumn{1}{c}{\hphantom{\fper}}&\multicolumn{1}{c}{\hphantom{\fper}}&\multicolumn{1}{c}{\hphantom{\fper}}&\multicolumn{1}{c}{\hphantom{\fper}}&\multicolumn{1}{c}{\hphantom{\fper}}&\multicolumn{1}{c}{\hphantom{\fper}}&\multicolumn{1}{c}{\hphantom{\fper}}&\multicolumn{1}{c}{\hphantom{\fper}}&\multicolumn{1}{c}{\hphantom{\fper}}&\multicolumn{1}{c}{\hphantom{\fper}}&\multicolumn{1}{c}{\hphantom{\fper}} &\multicolumn{1}{c|}{\hphantom{\fper}}\\\hline
Don't define as standard defaults    &0 &1 & \pv & \pv & & \p & & \splitYn &\cellcolor{cyan!50} &\cellcolor{cyan!50} &\cellcolor{cyan!50} &\cellcolor{cyan!50} &\cellcolor{cyan!50} &\cellcolor{cyan!50} &\cellcolor{cyan!50} \\
Different to login password    &0 &1  & & & & \pv & & \Y &\cellcolor{cyan!50} &\cellcolor{cyan!50} &\cellcolor{cyan!50} &\cellcolor{cyan!50} &\cellcolor{cyan!50} &\cellcolor{cyan!50} & \cellcolor{cyan!50}\\\hline

\multicolumn{1}{l}{\textbf{Password auditing}}&   & \multicolumn{1}{c}{}& \multicolumn{12}{c}{} &\multicolumn{1}{c|}{} \\\hline
Attempt to crack passwords     &0 &1 & \pper & \pper & \pper & \pper & \pper & \Y & \pv & \pv & & \f & & \pv & \Y\\\hline

\multicolumn{1}{l}{\textbf{Policies}}&   & \multicolumn{1}{c}{}& \multicolumn{12}{c}{} &\multicolumn{1}{c|}{} \\\hline
Establish clear policies &0 &2   & \pv & \pper & & \pv & & \Y &\cellcolor{cyan!50} &\cellcolor{cyan!50} &\cellcolor{cyan!50} &\cellcolor{cyan!50} &\cellcolor{cyan!50} &\cellcolor{cyan!50} & \cellcolor{cyan!50}\\\hline

\multicolumn{1}{l}{\textbf{Shoulder surfing}}&   & \multicolumn{1}{c}{}& \multicolumn{12}{c}{} &\multicolumn{1}{c|}{} \\\hline
Offer to display password   &0 &1  & & \pv & & \p & & \n & & & & & & & \Y\\
Enter your password discretely &0 &2 &\cellcolor{cyan!50} &\cellcolor{cyan!50}{\p} &\cellcolor{cyan!50} &\cellcolor{cyan!50} &\cellcolor{cyan!50} & \cellcolor{cyan!50}& \p & & & & & \p & \Y\\\hline

\multicolumn{1}{l}{\textbf{Storage}}&   & \multicolumn{1}{c}{}& \multicolumn{12}{c}{} &\multicolumn{1}{c|}{} \\\hline
Encrypt password files    &0 &1 & & \p & & \p & \p & \Y &\cellcolor{cyan!50} &\cellcolor{cyan!50} &\cellcolor{cyan!50} &\cellcolor{cyan!50} &\cellcolor{cyan!50} &\cellcolor{cyan!50} &\cellcolor{cyan!50} \\
Restrict access to password files    &0 &2 &  &  & & \p & \pv & \Y &\cellcolor{cyan!50} &\cellcolor{cyan!50} &\cellcolor{cyan!50} &\cellcolor{cyan!50} &\cellcolor{cyan!50} &\cellcolor{cyan!50} &\cellcolor{cyan!50} \\
Hash and salt passwords &0 &4 & & & \pv & \p & \p & \Y  &\cellcolor{cyan!50} &\cellcolor{cyan!50} &\cellcolor{cyan!50} &\cellcolor{cyan!50} &\cellcolor{cyan!50} &\cellcolor{cyan!50} & \cellcolor{cyan!50}\\
Encrypt passwords   &*4& 7 & \pv & & \pv & \p & \pv & \Y  &\cellcolor{cyan!50} &\cellcolor{cyan!50} &\cellcolor{cyan!50} &\cellcolor{cyan!50} &\cellcolor{cyan!50} &\cellcolor{cyan!50} &\cellcolor{cyan!50} \\
Don't hardcode passwords   &0 &1  & \p & \pper & \p & \p & & \Y  &\cellcolor{cyan!50} &\cellcolor{cyan!50} &\cellcolor{cyan!50} &\cellcolor{cyan!50} &\cellcolor{cyan!50} &\cellcolor{cyan!50} &\cellcolor{cyan!50} \\\hline

\multicolumn{1}{l}{\textbf{Throttling}}&   & \multicolumn{1}{c}{}& \multicolumn{12}{c}{} &\multicolumn{1}{c|}{} \\\hline
Throttle password guesses    &0 &8   & \pper & \pper & \pper & \p & & \Y & &  & & & & & \Y\\\hline

\multicolumn{1}{l}{\textbf{Transmitting passwords}}&   & \multicolumn{1}{c}{}& \multicolumn{12}{c}{} &\multicolumn{1}{c|}{} \\\hline
Don't transmit in cleartext    &0 &4 & \pperv & \pv & & \p & \pv & \Y &\cellcolor{cyan!50} &\cellcolor{cyan!50} &\cellcolor{cyan!50} &\cellcolor{cyan!50} &\cellcolor{cyan!50} &\cellcolor{cyan!50} & \cellcolor{cyan!50}\\
Request over a protected channel &0&2  & \pper & \pper & \p & \p & & \splitYn &\cellcolor{cyan!50} &\cellcolor{cyan!50} &\cellcolor{cyan!50} &\cellcolor{cyan!50} &\cellcolor{cyan!50} &\cellcolor{cyan!50} &\cellcolor{cyan!50} \\\hline

\multicolumn{1}{l}{}&   & \multicolumn{1}{c}{}& \multicolumn{12}{c}{} &\multicolumn{1}{c|}{} \\\hline
\textcolor{gray}{Don't allow users to paste passwords} &&   & \fperv & \pper & & \p & & \X & & \plogv & & & & \plogv & \X\\\hline

\captionsetup{width=0.7\linewidth}
\caption*{\f, filled circle: major cost.\hfill \p, empty circle: minor cost.  \hfill +, plus: positive cost.\newline
            $\colourone\colorboxblue^{@}$, superscript $\colourone@$: cost occurs at each login.  $\colourtwo\colorboxpink^{\sun}$, superscript $\colourtwo\sun$: cost occurs periodically.\newline
            \underline\colorboxgrey, underline: implies that variance existed in the costs that respondents indicated.\newline
            \splitYn : approval split between Yes and Neutral.
            }
\label{coststableendpageref}
\end{longtable}
}

\normalsize

 \section{Costs survey: results}\label{sec:cost-results}
 We received 44 participants for our end-user survey and 37 participants for our administrator survey. We received insightful comments about the costs from both a user and administrator perspective. A minimum of  eight end-users indicated the costs they associated with each piece of advice and a minimum of six administrators indicated the organisation costs they perceived, again for each piece of advice. Tab.~\ref{tab:table_of_costs} on Page~\pageref{tab:table_of_costs}~--~\pageref{coststableendpageref}  details the costs that users and administrators identified for each advice statement. In this section, we highlight the following: 1. the feedback we received from users and administrators on our suggested categories of costs, 2. the costs participants assigned to the advice and 3. participants' approval of the advice. 
 
\subsection{Cost category comments}
\subsubsection{User cost categories} After users had attempted to assign cost categories to the advice statements, they were asked at the end of the survey whether they agree with the cost categories that were used in the survey. 

For the user survey the prevailing answer was between `Somewhat' and `Yes': 21 end-users said Somewhat, 18 said Yes and 1 said No. The user who chose `No' said: ``In most cases the categories seemed not applicable to the points so most were of no cost to me.''
This is a valid point, for many advice statements there was only one cost category that users deemed to apply and for some advice such as ``Every user in an organisation must have their own account'', the majority of participants decided that there was no associated user costs. 

After asking participants whether they agreed with the cost categories, we asked whether there were any cost categories that they think should be added or removed. We received the following suggestions. One person said we could remove computing power. One person said a cost could be that the advice ``decreases sense of security in passwords''. Another participant suggested maybe including the cost category ``makes it harder to follow other advice''. They gave the example that needing to have long passwords would make it more difficult to not write them down somewhere. One participant suggested that there is a \textbf{``cost to personal stress related to constantly interacting with devices that require various different logins and passwords and eat up time and energy!!''.} All 34 other participants suggested no changes to the cost categories.

We did consider removing the user computing power cost but in the end decided that in some areas it might be relevant. For example for the the pieces of advice ``Keep anti-virus updated'' and ``Keep software updated'' the majority of users said that computing power was a minor cost. This can relate to a slow down of computing processes during an installation or waiting for restarts during updates. 

The ability of a participant to follow multiple pieces of advice simultaneously is important to consider. For example, Flor\^encio et al. show that never reusing a password and also random password choice is an impossible task outside the bounds of human memory~\cite{florencio2014password}. In fact, many respondents in our survey made the point that the advice ``Never reuse a password'' is impossible to uphold unless it is coupled with the use of a password manager. For this reason, to really get a sense of the value or effect of a piece of advice it is important to consider it in light of a complete security policy. 

A yes/no answer system would have made the survey easier for participants to understand and simplify the completion. We were eager to get information about severity and frequency but in retrospect a simplified version of the survey should have been considered. We did also notice that their was misunderstanding among participants about the meaning of the `Positive costs'. We think a different term should have been used to describe these that was more self-descriptive. Based on the answers, we believe some participants indicated `positive' when they were `positive' there was a cost there. These participants were always the minority so largely did not affect the results. 

\textbf{Finally, the stress associated with current password systems is exemplified by the respondents' encouragement for ``costs to personal health'' to be included as a category.} When we consider a security system and speak about usability, it is mitigating this stress and pressure that users are burdened by that we want to achieve. We believe this stress is a result of the human effort and mental strain that is encompassed within the existing categories. 

\subsubsection{Organisation cost categories}
As with users, after administrators had attempted to assign organisation costs to the advice statements, they were asked at the end of the survey whether they agree with the cost categories that were used in the survey. The majority of administrators agreed with the five cost categories that were used to denote organisation/administration costs in this survey. However, nearly as many said they `Somewhat' agreed. 13 administrators said Yes, 12 said Somewhat and 4 said No.

The most common comment we received was that user burdens were not included as a cost category. It is likely that administrators were not aware that a second survey existed aimed at users and the burdens they experience. However, \textbf{the fact that many administrators insisted that user burdens must be taken into account is reassuring. It shows a changing in the ethos within security development} and indicates that there could exist a changing perspective which is no longer viewing users as the enemy \cite{adams1999users}.

One respondent commented that the `resources' cost was not specific enough. From feedback we received during the survey, we acknowledged that a misunderstanding existed with the term `additional resources needed'. Some respondents viewed resources as additional personnel needed whereas we were referring to resources as physical purchases required. We categorised a need for additional personnel under increased help desk/user support time and/or time take to implement. We did include a clarification for this partway through the study stating that ``Resources refers to physical resources that may need to be acquired or purchased''. After this clarification the responses seemed more aligned with this definition. For this reason though, a resource cost is sometimes identified where it does not seem clear to us why it should be. We do leave it in as significant increases in necessary personnel and other forms of resources are valid forms of burdens for an organisation.

One participants mentioned that ``Everything has a cost sometimes it is small, but in many areas questioned in this survey, the benefit outweighs the cost''. \textbf{This indicates that this comparison of costs versus benefits of security advice is something that security administrators are required to do mentally for each policy consideration. }

Two participant mentioned other costs. One said that ``Some things need more than just help desk resources - development time, administration, auditing''. The second participant said ``There is more than just training and help support in terms of cost. Their is also engineering cost, audit cost, risk assessment cost, employee quality costs (not all employees can implement the policies discussed here in an enterprise environment). Technology cost. Enforcement cost. Incident response cost (for policy violations)''. These are all interesting areas for consideration. We would have viewed all of these as ``organisation time taken to implement the policy'' though employee quality costs is not something we considered.

Finally, one respondent mentioned that the extent of costs can depend strongly on the ethos of a company. They state that ``Having a business leadership team that fully supports IT Security and is prepared to champion it will for example make the initial and ongoing ``cost'' in terms of resource much easier. Many of the things here will depend not just on technology but the culture and maturity of the organisation''. This alludes to feedback we received throughout the survey and directly from some respondents: \textbf{the costs experienced differ for each organisation. While the general costs felt might be similar, the extent to which they are felt and the difficulty involved in implementing advice will strongly depend on the ethos size and sector or services of the company.} This model only provides a broad indication of the categories of costs felt by the majority.

\subsection{Cost of security advice}\label{sec:disc_costs} 
The costs associated with implementing and following advice are shown in Tab.~\ref{tab:table_of_costs}. In this table the most commonly used user cost category was \textit{increased risk of forgetting} and the most commonly used organisation cost category was \textit{user education required}. It is important to note that while organisation time is required to develop and run user education, user time is also necessary. Therefore, each time we see a \textit{user education} cost, we recall that this involves another burden for users. 

There were 200 costs identified across all the pieces of advice for the organisation and 109 costs identified for users. Of these the users identified that 28\% of them were major costs. Administrators identified that 16\% of the costs to the organisation were major. 37.5\% of these major organisation costs were attributed to user education.

Of the 200 organisation costs, 44\% were repeating costs, i.e. they were felt at each login or periodically. Of the users' 109 costs 46\% were repeating. However, nearly half (49\%) of the repeating organisation costs were attributed to user education, a cost that will be felt by users as well. Similarly, of the user repeating costs 40\% were related to the cost of forgetting as many attributed this as occurring at login. The cost of forgetting is a cost that is felt by the organisation as well since the users' only solution is often to contact the organisation help desk.

Of our eleven cost categories, help desk time, time to implement, user education, and all the user costs excluding computing power, involve a human directly. In these categories, advice is adding to the emotional burden or workload of the individuals involved. \textbf{Human are impacted by 87\% of all the costs identified. In the user survey we witness strong emotional responses towards the advice we asked users to analyse. We received responses such as ``I am sick of passwords and logins and they are making me less productive as I have to look up the passwords so often!!''.}

Users only agreed on three pieces of advice which have a positive impact on a cost category. These are: ``You should use a password manager to store your passwords'', ``Write your password down safely'' and ``You should alter your password and then reuse it on other websites or computers''. All three of these had a positive impact on \textit{memorability}.

\subsection{Approval of advice}
For each piece of advice, we asked users and administrators to indicate whether they approved of the advice, felt neutral about it, or disagreed with it. In the 7\textsuperscript{th} and 14\textsuperscript{th} columns of Tab.~\ref{tab:table_of_costs} we indicate the result that represented the reactions of the majority of respondents in each category. 

\subsubsection{Mutual consensus}
While we took the majority answer as the mutual consensus for  Tab.~\ref{tab:table_of_costs}, it is important to note that opinions on advice was divided. We found that for 32\% of the advice, there was at least one administrator respondent who disagreed with others on whether advice was valuable. More starkly, for 71\% of the advice considered by end-users, there was at least one user-respondent with a different opinions to others. While a 32\% difference could be put down to differences of opinion on practice or different interpretation of the advice statement, the 71\% disagreement among end-users shows a significant lack of consensus. Clearly, more user research or user education is needed to understand this.

The advice users were most conflicted about was:
\begin{itemize}
    \item Never reuse a password (19 approve, 13 disapprove, 9 neutral)
    \item Don't allow users to paste passwords (14 approve, 17 disapprove, 10 neutral)
\end{itemize}

The advice administrators were most conflicted about was:
\begin{itemize}
    \item Don't open emails from strangers (4 approve, 3 disapprove, 0 neutral)
    \item Include specific characters in your \underline{username} (3 approve, 2 disapprove, 2 neutral)
\end{itemize}

Tab.~\ref{tab:differing_u_ad} shows other examples of users or administrators disagreeing. Note that though the median number of administrator respondents to each sub-survey was 7, and the median number of user respondents to each sub-survey was 8, because users were only consulted about 56 pieces of advice to administrators' 68, we included some overlap in the user's surveys. This is why, for some pieces of advice, we had up to 41 respondents to a single question.

\renewcommand{\tabcolsep}{5pt}{
\begin{table}[]
    \centering
    \begin{tabular}{|c|c|c|c|c|c|c|c|c|}\hline
    \multicolumn{4}{|l|}{Administrators} &&  \multicolumn{4}{l|}{Users}\\\hline
    \rot{Decision}  &\rot{\#Approve }& \rot{\#neutral\hphantom{v}}& \rot{\#Disagree } & \textbf{Advice} & \rot{\#Approve}& \rot{\#neutral} & \rot{\#Disagree} &\rot{Decision} \\\hline
    \Y & \cellcolor{green}6 & 1 & 0     &  Never reuse passwords &  \cellcolor{green}19 & 9 & 13 & \Y\\\hline 
    \X & 2 & 0 & \cellcolor{red}6     &  Don't allow users to paste passwords &  14 &10 & \cellcolor{red}17 & \X\\\hline 
    \Y & \cellcolor{green}3 & 2 & 1     &  Password hints should not be stored &  3 & 2 & \cellcolor{red}4 & \X\\\hline
    \X & 0 & 2 & \cellcolor{red}6     &  Manually type URLs &  2 & \cellcolor{yellow}4 & 2 & \n \\\hline
    \Y & \cellcolor{green}7 & 0 & 0     &  When using a password manager create long random password &  3 & \cellcolor{yellow}4 & 2 & \n\\\hline
    \splitYn & \cellcolor{green}3 & \cellcolor{yellow}3 & 1     &  Don't choose ``remember me'' on your computer &  2 & \cellcolor{yellow}6 & 3 & \n\\\hline
    \Y & \cellcolor{green}6 & 1 & 0     &  Use two-factor authentication &  \cellcolor{green}32 & 15 & 13 & \Y\\\hline
    \Y & \cellcolor{green}4 & 0 & 3     &  Don't open emails from strangers &  \cellcolor{green}7 & 2 & 1 & \Y\\\hline
    \Y & \cellcolor{green}3 & 2 & 2     &  Include specific characters in your \underline{username} &  0 & 2 & \cellcolor{red}6 & \X\\\hline
    \Y & \cellcolor{green}4 & 2 & 2     &  Don't use dictionary words in your password &  2 & \cellcolor{yellow}5 & 1 & \n\\\hline
    \X & 2 & 1 & \cellcolor{red}4     &  Change your password regularly &  10 & 10 & \cellcolor{red}20 & \X\\\hline
    \Y & \cellcolor{green}4 & 1 & 1 & Alter and reuse passwords & 2&2&\cellcolor{red}5 &\X \\\hline
    \end{tabular}
    \caption{Differing opinions among user respondents and among administrator respondents. The majority rating  for each group is highlighted and indicated in the decision column.}
    \label{tab:differing_u_ad}\vspace{-2em}
\end{table}}

Tab.~\ref{tab:differing_u_ad} shows differing opinions among user respondents and among administrator respondents. The majority rating  for each group is highlighted and indicated in the decision column. Notice that in some places, the users and administrators come to a different consensus about whether a piece of advice is valuable. For example, users believe they should be allowed to store password hints, but administrators disagree. 

Users were conflicted about whether typing URLs was good advice, however, administrators unanimously (excluding neutral responses) agreed it was bad advice.  Administrator respondents said ``It might be nice in theory, but impossible to implement and enforce in practice'' and ``There may be occasions when this is relevant but typo-squatting also poses a risk besides phishing''. This advice might have security benefits, but it introduces a new attack vector and also results in a significant time loss for anyone who follows it.  

The advice ``Include specific characters in your username'' divided administrator opinions and users strongly disagreed with it. In a 2007 research paper~\cite{florencio2007strong} Flor{\^e}ncio et al. determined that rather than making the secret password more complex, complexity could instead be added to the username. Users can choose and record a long username and this will likely offer effective protection from a bulk guessing attack. The very attack that a complex or long password is hoping to achieve. One administrator disapproving of this advice said ``good for passwords, over complicates usernames as password should be secure, usernames are often generated based on firstname.lastname etc''. One end-user respondent commented on the advice said ``Stupid and pointless'' and another said ``Can't see a good reason for this, but with a password manager it wouldn't be too painful''. If this policy was being implemented for an organisation then these responses emphasise the need for it to be coupled with effective user-education and validation.

Administrators were generally in favour with altering a password before reusing it at another site. But most users surveyed did not approve of the advice. One user who disagreed with the advice said ``Impossible to remember what password goes with what site''. A different user who disagreed said ``Any method to your password creation makes it less secure''. One administrators commented that ensuring alterations before password reuse should be a strict policy that is reinforced by user education.


Users, were torn on the advice stating that it should not be possible to paste your password when logging in. The comments given emphasise the differing opinions. One user said: ``If someone is trying to copy/paste a password, they're probably beyond help''. While another user said: ``This is horrendous advice that leads to problems using password managers. It encourages using crappy passwords.''

Throughout the survey, many users showed a willingness to follow advice provided it seemed to have a benefit. For example, for the advice regularly change passwords one user commented: ``Undoubtedly sensible, undoubtedly annoying''. There is substantial evidence in the responses~\cite{hazel_github} that user education has been effective. User echoed the advice that was given in historic security documents \cite{nist2003/6}. However, few users and administrators showed an awareness of new recommendation which have superseded the legacy advice~\cite{nist2017}. For example, the 2017 NIST advice recommends not introducing restrictions on the characters allowed in passwords. Yet most users and administrators indicated that they agreed with forcing the use of certain characters in passwords. Note that this new NIST advice have been available and circulating since 2017 and our user survey was completed in October 2020.

Overall, understanding users' approval of advice is a valuable stepping stone to improved user education and mutual respect between users and policy enforcers. As Adams and Sasse show, users will simply employ less secure workarounds if security policies don't allow them to do their job effectively~\cite{adams1999users}.

\section{Benefits model} \label{sec:benefits}

Now we ask the question: What are the benefits of this advice? From our analysis of costs, it is evident that advice is generally not attempting to make authentication more usable. Instead, advice is attempting to guide the creation of robust authentication systems with the goal of preventing unauthorized access to an account or to data~\cite{arnell2012systematic,florencio2010security,herley2009so,shay2009comprehensive}. This is what we define as the goal of the advice. Therefore, the benefit of a piece of advice is the affect it has towards achieving this goal. In this section, we give a method for identifying what benefits a single piece of advice can bring.

\subsection{Benefit categories}

Unauthorized access has many avenues and one piece of advice is unlikely to protect against them all. For example, requiring passwords of length 12 might help against a password guessing attack, but will do nothing to protect against a phishing attack. In order to represent how `beneficial' a piece of advice is, we would like to know in which ways the piece of advice protects against unauthorized access, if at all.

To this end, we require a concise list of the different methods an adversary could use to gain access to a protected account. The NIST 2017 Digital Identity Guidelines \cite{nist2017} includes a table of ``Authenticator Threats''. Given that this NIST 2017 authentication document has been a highly regarded \cite{nistcox} and thoroughly researched document we accept these as a concise list of the attack vectors against authentication. Below, in our Tab.~\ref{tab:attacks}, we list the attacks that NIST identified (Tab.~8-1 in~\cite{nist2017}). Included is the name, details and example for each different attack type. In places we diverge slightly from the official description and/or example provided in the NIST document and instead indicate our interpretation of the threat.

\begin{table}[h]
    \small
    \centering
      \caption{\label{tab:attacks} Attack types on authentication}
      \renewcommand{\arraystretch}{0.9}
\begin{tabular}{p{0.15\linewidth}|p{0.39\linewidth}|p{0.39\linewidth}}
    Attack type                             & Description                       & Examples\\\hline
    Assertion manufacture or modification   &  The assertion is used to communicate the result of the authentication process from the verifier to the Relying Party (RP). The verifier and the RP may be the same entity, or separate entities. & Use of assertion replay to impersonate a valid user or leakage of assertion information through the browser \cite{somorovsky2012breaking}.\\\hline
     Physical theft         & A physical authenticator or physical device used in the authentication process is stolen by an attacker.  &A hardware cryptographic key (e.g. a USB authenticator), a phone, computer or one-time-password device is stolen.\\\hline
     Duplication            & The subscriber’s authenticator has been copied with or without their knowledge. & Passwords or private key written on paper or stored in an electronic file are copied. A counterfeit biometric authenticator is manufactured.\\\hline
     Eavesdropping          & The authenticator secret or authenticator output is revealed to the attacker as the subscriber is authenticating. &  Passwords are physically observed during keyboard entry or intercepted by keystroke logging software or recorded during transmission or via network packet sniffing.\\\hline
     Offline Cracking       & An offline guessing attack is an analytical guessing attack by an attacker, it requires little to no communication with the system under attack. & A dataset of passwords or keys which are hashed \& salted or encrypted are made available to an attacker (leaked). Using a dictionary or brute force guessing method the attacker attempts to guess the plaintext values of these protected secrets.\\\hline
     Side Channel Attack    & This attack leverages an aspect of the implementation of the computer system or security device. &  An attacker exploits information about a cryptographic key gathered from power, timing or audio data.\\\hline
     Phishing or Pharming   & Fooling the subscriber into thinking the attacker is a verifier. Phishing: electronic communication masquerading as the verifier. Pharming: directing a website's traffic to a masquerading fake site. & Phishing: A password or key is revealed by a bank subscriber in response to an email inquiry from a phisher pretending to represent the bank. Pharming: a password is revealed by the subscriber at a fake verifier website reached through DNS spoofing \cite{chaudhry2016phishing} \cite{APWG2019report}\\\hline
     Social Engineering     & The attacker establishes a level of trust with a subscriber in order to convince the subscriber to reveal their authenticator secret or authenticator output. & An attacker masquerading as a system administrator makes a telephone inquiry requesting the victim's password.\\\hline
     Online Guessing        & The attacker connects to the online server of the verifier and attempts to guess the valid authenticator output for one or multiple users. & An attacker who knows the usernames for all the accounts guesses the top ten most popular passwords in order to try to access them. An attacker who knows neither the username nor password guesses combination of both to try to unlock the account.\\\hline
     Endpoint Compromise    & Malicious code on the endpoint authenticates without the victim's consent, compromises the authenticator or causes authentication to other than the intended verifier. & Malicious code can steal data from the users' device. Malicious code can be used to conduct man-in-the-middle attacks on all connections. New trusted certificates can be installed on the users' device.  \\\hline
     Unauthorized binding   & An attacker is able to cause an authenticator under their control to be bound to a subscriber’s account.& Forcing a password reset to change password to one the attacker knows. Or creation of a USB authenticator and linking it to a users' account. \\\hline
\end{tabular}
\end{table}
     \normalsize

\subsection{Representation of advice benefits} 
As with costs, which could be larger or smaller, the benefits provided by different advice also varies. The benefits of advice therefore should reflect the probability of each attack being successful. Given some baseline chance of each attack, we reflect on whether a piece of advice increases or decreases the chance that that attack is successful. We show the result of this analysis for the 79 advice statements in Tab.~\ref{tab:table_of_benefits}, page \pageref{tab:table_of_benefits}--\pageref{endofbenefitstable}.

We use the symbols \hphantom{l}\veryup \hphantom{l}and \hphantom{l}\verydown \hphantom{l} to indicate an increase or decrease respectively in the probability of success for the attack type. \hphantom{l}\up  \hphantom{l}and \hphantom{l}\down  \hphantom{l}indicate less significant increases or decreases in the probabilities of success for the attacks. An underline, \underline{\colorboxgrey}, indicates that the improvement is not directly enforceable. This could be because the advice is too vague and therefore there is ambiguity on how an organization or user might follow it. Or simply that it is impossible for an organization to bind their users to following the advice.

In supplementary material we include all explanations for why we believe the chances of certain attacks are impacted by the advice~\cite{hazel_github}. It is worth noting that when determining this assignment of benefits, we took input from stakeholders and experts including giving presentations at conferences/workshops to gather insight from specialists (\cite{usenix17_posters, murray2017evaluating, Passwordscon_talk,HEAnet_talk}), statistics from our university IT department and insights from our University's information security manager. 
  
\subsection{Baseline for comparison}
Throughout our table we mark each piece of advice as either increasing or decreasing the chance of compromise. However, there is an interesting question about what baseline the advice increases or decreases the chance of compromise relative to. We describe two mechanisms for determining this baseline.

We first thought of the baseline as being a passive version of the advice. For example, if the advice is: ``Email up-to-date and secure'' or ``Encrypt passwords'' then the baseline we measure the improvements from is doing nothing. This made sense for a lot of the advice statements, but not all. For example, the advice ``Do not store hints'' is already a passive action, but we do not want to measure its affects against itself.

Our second consideration for the baseline was the opposite of what the advice stated. For example, for the advice ``do not store hints'' we consider the difference in attack threat between this and ``do store hints''. Or maybe less strictly, ``you are allowed to store hints''.

We have chosen to use this `opposite' method as it appears the logical comparison for most pieces of advice. The downside is that the opposite can, in a few cases, be too `strict'. For example, the opposite of ``do not include names'' and ``don't repeat characters'' are ``include names'' and ``repeat characters'' respectively, both unnatural pieces of advice. Two pieces of advice had ambiguous opposites. The opposite of ``write down safely'' could either be ``don't write down'' or ``write down, but not safely''. In this case, we chose to consider the opposite as ``don't write down''. Similarly, we choose the opposite of ``alter and reuse passwords'' to be that reuse is allowed even without altering.

As said, for our purposes in this paper we identify the increase or decrease in the chance of attack by comparing to the opposite of the specific piece of advice. However, in a real-world situation a proposed policy could simply be compared in light of the current/previous policy in place.

No matter what baseline method is chosen, the discussions provided in our supplementary material~\cite{hazel_github} should be relevant when assessing benefits.

\thispagestyle{empty}

\renewcommand{\tabcolsep}{8pt}{
\begin{longtable}{lrr||c|c|c||c|c|c||c|c|c||c|c|c|} 
\caption{\label{tab:table_of_benefits}Benefits of implementing password advice}
\\\cline{4-14}
\multicolumn{1}{c}{} &\multicolumn{2}{c}{} & \multicolumn{11}{|c|}{\rule{0pt}{1em}Attack Types}\\\cline{4-14}
\textit{Advice to users}&\rot{\#\ sources contradicting} & \rot{\#\ sources advising}& \rot{\shortstack{Assertion manufacture \\or Modification}} & \rot{\shortstack{Physical Theft}} & \rot{\shortstack{Duplication}} & \rot{\shortstack{Eavesdropping}} &  \rot{\shortstack{Offline Guessing Attacks\hphantom{W}}} & \rot{\shortstack{Side Channel Attack}} & \rot{\shortstack{Phishing or Pharming}}  & \rot{\shortstack{Social Engineering}} & \rot{\shortstack{Online Guessing}} & \rot{\shortstack{Endpoint Compromise}}
           & \rot{\shortstack{Unauthorized binding}}   \\\hline
\endfirsthead 
\cline{4-14}
\multicolumn{1}{c}{} &\multicolumn{2}{c}{} & \multicolumn{11}{|c|}{\rule{0pt}{1em}Attack Types}\\\cline{4-14}
&\rot{\#\ sources contradicting} & \rot{\#\ sources advising}& \rot{\shortstack{Assertion manufacture \\or Modification}} & \rot{\shortstack{Physical Theft}} & \rot{\shortstack{Duplication}} & \rot{\shortstack{Eavesdropping}} &  \rot{\shortstack{Offline Guessing Attacks\hphantom{W}}} & \rot{\shortstack{Side Channel Attack}} & \rot{\shortstack{Phishing or Pharming}}  & \rot{\shortstack{Social Engineering}} & \rot{\shortstack{Online Guessing}} & \rot{\shortstack{Endpoint Compromise}}
           & \rot{\shortstack{Unauthorized binding}}   \\\hline
\endhead

\multicolumn{1}{l}{\textbf{Backup password options}}     &   & \multicolumn{1}{c}{} & \multicolumn{10}{c}{} &\multicolumn{1}{c|}{} \\\hline
Email up-to-date and secure         &0 & 3          &           &        &           &    \underline\down       &                  &           &    \underline\down       &         &   \underline\down           &   \underline\down  &\\
Security answers difficult to guess &0 & 3          &           &        &           &            &                  &           &            &         &   \underline\down           &      &\\
Do not store hints                  & 0 & 2         &           &        &           &            &    \underline\verydown     &           &            &    \underline\down     &   \underline\verydown   &      &\\\hline

\multicolumn{1}{l}{\textbf{Composition}}     &   & \multicolumn{1}{c}{} & \multicolumn{10}{c}{} &\multicolumn{1}{c|}{} \\\hline
Must include special characters     & 5 &  7  &      &           &           &           &   \verydown           &           &           &           &  \verydown        &      &\\
Don't repeat characters             & 0 &  3  &      &           &           &           &   \verydown           &           &           &           &  \verydown        &      &\\
Enforce restrictions on characters  & 1 & 12  &      &           &           &           &   \down            &           &           &           &  \down        &      &\\\hline

\multicolumn{1}{l}{\textbf{Keep your account safe}}     &   & \multicolumn{1}{c}{} & \multicolumn{10}{c}{} &\multicolumn{1}{c|}{} \\\hline
Check web pages for SSL/TLS             &0 &1    &       &            &             &    \underline\verydown  &      &           &     \underline\down   &              &  &      &\\
Manually type URLs                  &0 & 1   &       &            &  \underline\veryup   &               &      &           &     \underline\verydown   &              &  &      &\\
Don't open emails from strangers    & 0 & 1  &       &            &             &               &      &           &     \underline\verydown   &  \underline\verydown   &  &      &\\
Keep software updated               &0 & 2   &       &            &             &    \underline\verydown  &      &  \underline\verydown&                 &              &  & \underline\verydown   &\\
Keep anti virus updated             &0 & 2   &       &            &             &               &      &            &                 &              &  & \underline\verydown   &\\   
Log out of public computers         &0 & 2   &       &            &             &        \underline\verydown         &      &            &               &              &  &    &\\
Password protect your phone.        &0 & 1  &       &            &             &               &      &            &                 &              &  & \underline\verydown   &\\\hline

\multicolumn{1}{l}{\textbf{Length}} & & \multicolumn{1}{c|}{} & \multicolumn{10}{c}{} &\multicolumn{1}{c|}{} \\\hline
Minimum password length                 &0 &13  &   &   &   &  &  \verydown   &   &   &   &  \verydown  &  &  \\
Enforce maximum length (\textless40)    &1 &3   &   &   &   & \veryup & \veryup & &   &   &  \veryup &&\\\hline

\multicolumn{1}{l}{\textbf{Password managers}}     &   & \multicolumn{1}{c}{} & \multicolumn{10}{c}{} &\multicolumn{1}{c|}{} \\\hline
Use a password manager         &1 &2      &       &            &    \underline\veryup    &         &   \underline\verydown          &           &    \verydown   &       &        \underline\verydown        &      &\\
Create long random passwords    &0 &1     &       &            &                &         &   \underline\verydown &           &       &       &  \underline\verydown    &      &\\\hline

\multicolumn{1}{l}{\textbf{Personal Information}}     &   & \multicolumn{1}{c}{} & \multicolumn{10}{c}{} &\multicolumn{1}{c|}{} \\\hline
Don't include personal information  &1 & 5 &      &           &           &         &   \verydown     &           &           &           &  \verydown        &      &\\
Must not match account details      &0 & 8 &      &           &           &         &   \verydown     &           &           &           &  \verydown        &      &\\
Do not include names                &1 & 6 &      &           &           &         &   \verydown     &           &           &           &  \verydown        &      &\\\hline

\multicolumn{1}{l}{\textbf{Personal password storage}}     &   & \multicolumn{1}{c}{} & \multicolumn{10}{c}{} &\multicolumn{1}{c|}{} \\\hline
Don't leave in plain sight     &0 &4    &       &            &    \underline\verydown    &         &       &           &           &           &     &      &\\
Don't store in a computer file &1 &2    &       &            &     \underline\verydown   &         &       &           &           &           &     &      &\\
Write down safely              &1 &6    &       &     \underline\up       &   \underline\up      &         &       &           &           &           &     &      &\\
Don't choose "remember me"     &0 &3    &       &            &     \underline\verydown   &         &       &           &           &           &     &      &\\\hline

\multicolumn{1}{l}{\textbf{Phrases}}     &   & \multicolumn{1}{c}{} & \multicolumn{10}{c}{} &\multicolumn{1}{c|}{} \\\hline
Don't use patterns                 &0 & 6  &       &           &           &         &   \verydown     &           &           &           &  \verydown        &      &\\
Blocklist common passwords         &0 & 2  &       &           &           &         &   \verydown     &           &           &           &  \verydown        &      &\\
Take initials of a phrase          &0 & 4  &       &           &           &         &   \underline\verydown     &           &           &           &  \underline\verydown        &      &\\
Don't use published phrases        &1 & 2  &       &           &           &         &   \verydown     &           &           &           &  \verydown        &      &\\
Substitute symbols for letters     &1 & 2  &       &           &           &         &   \underline\verydown     &           &           &           &  \underline\verydown        &      &\\
Don't use words                    &0 & 16 &       &           &           &         &   \verydown     &           &           &           &  \verydown        &      &\\
Insert random numbers and symbols  &1 & 4  &       &           &           &         &   \underline\verydown     &           &           &           &  \underline\verydown        &      &\\\hline

\multicolumn{1}{l}{\textbf{Reuse}} & & \multicolumn{1}{c}{} & \multicolumn{10}{c}{} &\multicolumn{1}{c|}{} \\\hline
Never reuse a password          &*5 & 6  &      &           &           &           &      \underline\verydown           &         &   &   &     \underline\verydown      &       &\\\
Alter and reuse passwords       & 3 & 3  &      &           &           &           &      \underline\down           &         & \underline\up  &  \underline\up &     \underline\down      &       &\\
Don't reuse certain passwords   & 0 & 5  &      &           &           &           &      \underline\down           &         & \underline\up  &  \underline\up &     \underline\down      &       &\\\hline           

\multicolumn{1}{l}{\textbf{Sharing}}     &   & \multicolumn{1}{c}{} & \multicolumn{10}{c}{} &\multicolumn{1}{c|}{} \\\hline
Never share your password           &0 &9    &       &            &       &    \underline\verydown    &      &           &       &    \underline{\verydown}   &  &      &\\
Don't send passwords by email       &0 &3   &       &            &       &    \underline\verydown     &      &           &       &    \underline{\verydown}   &  &      &\\
Don't give passwords over phone     &0 &1   &       &            &       &    \underline\verydown     &      &           &       &    \underline{\verydown}   &  &      &\\\hline

\multicolumn{1}{l}{\textbf{Two factor authentication (2FA)}}        &   & \multicolumn{1}{c}{} & \multicolumn{10}{c}{} &\multicolumn{1}{c|}{} \\\hline
Use 2FA using app or special device    &0 &1    &      &   \underline\up    &        &       &       &       &   \verydown    &       &   \verydown   &      & \\
Use 2FA on phone                 &0 &1    &      &   \up    &       &  \up     &       &    \up   &   \verydown    &       &   \verydown   &   \up   & \\
Use 2FA for remote accounts            &0 &1    &      &    \underline\up    &       &  \veryup   &       &       &   \verydown    &       &   \verydown   &     & \\\hline

\multicolumn{1}{l}{\textbf{Username}}        &   & \multicolumn{1}{c}{} & \multicolumn{10}{c}{} &\multicolumn{1}{c|}{} \\\hline
Enforce composition restrictions    &0 &1    &      &       &       &       &       &       &       &       &   \verydown   &      & \\
Don't reuse username                &0 &1    &      &       &       &       &   \underline\down    &       &       &       &   \underline\verydown   &      & \\\hline
\multicolumn{14}{l}{}\\
\multicolumn{14}{l}{\textit{Advice to organisations}}
\\\hline
\multicolumn{1}{l}{\textbf{Administrator accounts}} & & \multicolumn{1}{c|}{} & \multicolumn{10}{c}{} &\multicolumn{1}{c|}{} \\\hline
Not for everyday use            &0&1    &       &           &      &   \verydown  &        & \verydown      &           &           &               &           &   \\ 
Must have it's own password     &0&2    &       &           &      &   \verydown  &        & \verydown      &           &           &               &           &   \\ 
Should have extra protection    &0&2    &   \underline{\down}    & \underline{\down}  & \underline{\down}  & \underline{\down}  &\underline{\down}    & \underline{\down}      &   \underline{\down}&     \underline{\down}  & \underline{\down}  & \underline{\down} &  \underline{\down} \\ \hline

\multicolumn{1}{l}{\textbf{Backup work}} & & \multicolumn{1}{c|}{} & \multicolumn{10}{c}{} &\multicolumn{1}{c|}{} \\\hline
Make digital \&\ physical back-ups.  &0 &1 &       &    \up   &      &     &        &       &           &          &        &           &   \\ \hline

\multicolumn{1}{l}{\textbf{Default passwords}} & & \multicolumn{1}{c|}{} & \multicolumn{10}{c}{} &\multicolumn{1}{c|}{} \\\hline
Change all default passwords.&0 &4     &       &           &      &     &        &           &           &           &  \verydown &           &   \\ \hline 

\multicolumn{1}{l}{\textbf{Expiry}} & & \multicolumn{1}{c|}{} & \multicolumn{10}{c}{} &\multicolumn{1}{c|}{} \\\hline
Store history to eliminate reuse &0 &5  &       & \veryup  &       &       & \verydown  &           &           &           & \verydown  &           &   \\
Change your password regularly  &4 &8   &       &           &       &       & \down  &           &           &           & \down  &           &   \\
Change if suspect compromise    &0 &10  &       &           &       &       & \verydown &           &           &           & \verydown  &           &   \\\hline

\multicolumn{1}{l}{\textbf{Generated passwords}} & & \multicolumn{1}{c|}{} & \multicolumn{10}{c}{} &\multicolumn{1}{c|}{} \\\hline
Use random bit generator        &*2 &2  &       &               &           &     &    \verydown    &       &           &          &   \verydown        &           &   \\ 
Must aid memory retention       &0 &2   &       &               &           &     &    \veryup     &       &           &          &   \veryup         &           &   \\ 
Must be issued immediately      &0 &1   &       &               & \verydown &     &                 &       &           &          &                    &           &   \\ 
Distribute in a sealed envelope &0 &1   &       &  \veryup     &  \up&     &    \verydown    &       &           &          &                    &           &   \\ 
Only valid for first login      &0 &1   &       &               &  \verydown&     &                 &       &           &          &                    &           &   \\\hline 

\multicolumn{1}{l}{\textbf{Individual accounts}} & & \multicolumn{1}{c|}{} & \multicolumn{10}{c}{} &\multicolumn{1}{c|}{} \\\hline
One account per user            &0 &4   &      &           &  \verydown    &     &        &           &  \verydown        &     \verydown      &   &    \verydown       & \verydown  \\ 
Each account password protected &0&3    &        &    &      & & \verydown   &\down   &          &           &\verydown&          & \\ \hline

\multicolumn{1}{l}{\textbf{Input}}& & \multicolumn{1}{c|}{} & \multicolumn{10}{c}{} &\multicolumn{1}{c|}{} \\\hline
Don't perform truncation  &0 &1 &       &           &       &       & \verydown  &           &           &    \verydown       & \verydown  &           &   \\ 
Accept all characters       &1 &1 &       &           &       &       & \verydown  &           &           &                    & \verydown  &  \underline{\up}    &   \\\hline
\multicolumn{1}{l}{\textbf{Keep system safe}} & & \multicolumn{1}{c|}{} & \multicolumn{10}{c}{} &\multicolumn{1}{c|}{} \\\hline
Implement Defense in Depth          &0 &2 &   \underline{\down}         &   \underline{\down}   & \underline{\down}         & \underline{\down}         &\underline{\down}    & \underline{\down}       &   \underline{\down}&     \underline{\down}  & \underline{\down}  & \underline{\down} &  \underline{\down} \\
Implement Technical Defenses        &0 &1 &   \underline{\down}         & \underline{\down}     & \underline{\down}         &   \underline{\down}       &\underline{\down}    & \underline{\down}       &   \underline{\down}&     \underline{\down}  & \underline{\down}  & \underline{\down} &  \underline{\down} \\
Apply access control systems        &0 &1 &   \underline{\verydown} &                   & \underline{\verydown} & \underline{\verydown}  &                  &  \underline{\down}    &  \underline{\down} &     \underline{\down}  & \underline{\down}  & \underline{\down} &  \underline{\down} \\
Monitor and analyse intrusions      &0 &1 & \underline{\verydown}               &   & \underline{\down}  &   &    &  \underline{\down}  &   &   & \underline{\verydown} &  &  \underline{\verydown} \\
Regularly apply security patches    &0 &1 &                         &   &   & \down  &    &  \down      &   &       &   & \verydown &  \down \\\hline

\multicolumn{1}{l}{\textbf{Network: Community strings}} & & \multicolumn{1}{c|}{} & \multicolumn{10}{c}{} &\multicolumn{1}{c|}{} \\\hline
\shortstack{Don't define as standard default}   &0 &1   &       &       &      &     &        &           &           &           &  \verydown &           &   \\ 
Different to login password                     &0 &1   &       &           &      &   \verydown  &        & \verydown      &           &           &               &           &   \\ \hline

\multicolumn{1}{l}{\textbf{Password auditing}} & & \multicolumn{1}{c|}{} & \multicolumn{10}{c}{} &\multicolumn{1}{c|}{} \\\hline
    Attempt to crack passwords      &0 &1  &       &           &      &     &        &       &           &           &         \verydown      &           &   \\ \hline

\multicolumn{1}{l}{\textbf{Policies}}& & \multicolumn{1}{c|}{} & \multicolumn{10}{c}{} &\multicolumn{1}{c|}{} \\\hline
Establish clear policies    &0 &2  &       &           &      &     &        &     &           &           &               &           &   \\ \hline

\multicolumn{1}{l}{\textbf{Shoulder surfing}} & & \multicolumn{1}{c|}{} & \multicolumn{10}{c}{} &\multicolumn{1}{c|}{} \\\hline
Offer to display password       &0 &1   &       &           &           &  \veryup &           &           &           &           &           &           &   \\
Enter your password discretely  &0 &2   &       &           &           &  \verydown &           &           &           &           &           &           &   \\\hline

\multicolumn{1}{l}{\textbf{Storage}} & & \multicolumn{1}{c|}{} & \multicolumn{10}{c}{} &\multicolumn{1}{c|}{} \\\hline
Encrypt password files                  &0 &1  &        &           & \verydown   &     &     \verydown   &       &           &           &            &           &  \verydown \\
Restrict access to password files       &0 &2 &       &    \verydown   &      &     &    \verydown    &       &           &           &        &           &  \verydown \\
Hash and salt passwords       &0 &4   &       &           &           &       & \verydown  &           &           &           &           &           &   \\
Encrypt passwords           &*4& 7  &       &           &           &       & \verydown &           &           &           &           &           &   \\
Don't hardcode passwords    &0 &1   &       &           &           &       & \verydown  &           &           &           &           &           & \\\hline

\multicolumn{1}{l}{\textbf{Throttling}} & & \multicolumn{1}{c|}{} & \multicolumn{10}{c}{} &\multicolumn{1}{c|}{} \\\hline
Throttle password guesses       &0 &8  &       &           &      &     &        &       &           &           &         \verydown      &           &   \\ \hline

\multicolumn{1}{l}{\textbf{Transmitting passwords}} & & \multicolumn{1}{c|}{} & \multicolumn{10}{c}{} &\multicolumn{1}{c|}{} \\\hline
Don't transmit in cleartext         &0 &4   &   \verydown    &           &      &   \verydown  &        &           &           &           &  \verydown &           &   \\ 
Request over a protected channel    &0&2    &   \verydown    &           &      &   \verydown  &        &           &           &           &  \verydown &           &   \\ \hline 

\multicolumn{1}{l}{}&   & \multicolumn{1}{c}{}& \multicolumn{10}{c}{} &\multicolumn{1}{c|}{} \\\hline
\textcolor{gray}{Don't allow users to paste passwords}&&    &       &           &      &     &        &           &           &           &  &           &   \\\hline

\caption*{\verydown \hphantom{w}Decreases the probability of attack.\hfill \veryup \hphantom{w}Increases the probability of attack.\newline
\up \hphantom{w}Minorly increases the probability of attack.\hfill
\down \hphantom{w}Minorly decreases the probability of attack.\newline
\underline{\colorboxgrey}, underline: advice is impossible to enforce; it must be followed voluntarily or in a certain way.
}
\end{longtable}}
\label{tab2:endofbenefitstable}
\label{endofbenefitstable}   \normalsize
\renewcommand{\arraystretch}{1}

\section{Benefits: Results}\label{disc_ben} 



Tab.~\ref{tab:table_of_benefits} starting on Page~\pageref{tab:table_of_benefits}~--~\pageref{tab2:endofbenefitstable} shows the benefits of implementing each of the 78 advice statements that we collected and the extra `canary' statement that we added in (don’t allow users to paste passwords). In this section, we will discuss the results from this assignment of benefits. First, we highlight an example assignment of benefits for the piece of advice on ``Password managers''. Then we fill discuss statistics on the types of attacks we protect against.  

\subsection{Benefits identified}
Overall, we identified 177 positive benefits (decreasing the chance of attack) and 26 negative benefits (increasing the chance of attack). 69\% of the negative benefits relate to the user advice and 31\% of the negative benefits relate the the organisation advice. A more even split of 43\% to 57\% respectively exists for the positive benefits. This means the organisation advice results in more benefits despite it also coming with lower costs.

\subsection{Example allocation of benefits}
Below, we take two advice categories and provide our discussion describing the benefits that were identified for each advice statement in that category. The two categories are password managers and generated passwords. For the discussion of each advice statement in Tab.~\ref{tab:table_of_benefits}, see~\cite{hazel_github}.

\subsubsection{Example Discussion: Password Manager}
\paragraph{Use a password manager}
 The security benefits of a password manager will depend heavily on both how it is utilized by the user and also on the capabilities on of the specific software that users are using. For this reason all benefits are dependent on the implementation.
 
 A password manager greatly reduces the users' memory load and, by extension, a user can then use passwords that are as long, random and complex password as they wish. Thus this act will increase security. A password manager does mean that the user is relying on an external agent to store their passwords and therefore if this agent is compromised or if the users password for this particular account is compromised then the passwords of all the users' accounts are compromised. Therefore we consider this to be a new way in which the users' passwords can be duplicated. 

Password managers that automatically fill in the users' credentials with no user interaction do have some corner case vulnerabilities \cite{silver2014password}. Though this same paper shows that a password manager can provide more security than the normal manual typing of the password. For example, password managers can be effective against phishing and pharming attacks.

\paragraph{Create long random passwords}
This piece of advice was given in the context of a password manager. ``Configure your password manager to create 30--50 random characters with a mixture of upper- and lower-case letters, numbers, and symbols.'' It has the same benefits as creating a complex long password but without the user memory costs.

\subsubsection{Example Discussion: Generated passwords}
\paragraph{Must be issued immediately}
This decreases the chance that generated passwords are stolen before they are told to the user. If passwords were created in advance they would likely be recorded as administrators could not remember multiple generated passwords. Therefore these passwords could be duplicated while in storage.

\paragraph{Distribute in a sealed envelope} 
This increases the chance that the password is physically stolen as the envelope could be taken. The password could also be duplicated since it has been recorded. If an adversary opens the envelope and duplicates the password then it will go undiscovered if the adversary places the password page in a new envelope and reseals it. The benefit of the sealed envelope is that an observational, audible or network eavesdropping attack is less likely. 

\paragraph{Only valid for first login} 
Because these generated passwords are often issued and created by administrators the user has no confidence in the security of their password up until the point they receive it. Maintaining a rule that passwords must be changed at first login means that the user can now have complete control over the security of this new password. This advice then protects against previous duplication of the password.

\subsection{Ambiguous advice}
We mentioned that an underline, \underline{\colorboxgrey}, in Tab.~\ref{tab:table_of_benefits} indicated that the improvement is not directly enforceable. We find that over half (52\%) of the advice we collected was either un-enforceable or too vague. This unenforceable advice was split evenly between the organisation and user advice statements. For example, ``Implement defense in depth'' can be divided into three categories: physical controls, technical controls and administrative controls. The security defense in depth can provide depends on exactly what strategies are deployed. They have the potential to mitigate any of the eleven attack types but without knowing what is implemented we cannot say exactly what the security advantages or disadvantages are.


\subsubsection{Negative benefits}
Observe that overall, the advice does seem to decrease the chance of compromise (most arrows point down, indicating a decrease in the chance of compromise). This should be unsurprising as this is after all, the aim of authentication advice. However, there are some areas where there are increases in the chance of attacks. Eight advice statements have major negative benefits and six pieces of advice have minor negative benefits. That is, in some areas, these pieces of advice can increase the probability of compromise. The remaining 65 advice statements all show improvements for security. 

\subsection{Frequency of attack protection}
In the current model framework, it might not always be meaningful to compare the security impact of one piece of advice against another. Take, for example, one piece of advice which decreases the probability of compromise against one attack type, and a piece of advice that decreases the probability of compromise against three attack types. It is likely that the latter piece of advice is ``better'' but in reality, different attacks occur with higher frequency than others and therefore protecting against one attack which occurs regularly might be more effective than protecting against three rare attack types.

This leads us to an interesting question on the frequency with which the different types of attacks are successful, and whether there is more advice against the more frequent attack types. Phishing, for example, occurs continuously, from targeted spear phishing attacks to mass phishing emails~\cite{anti2018phishing}. Whereas side channel attacks, while they attract interest from researchers, appear to have small real-world chance of occurring.

\begin{figure} 
\centering 
   \tcbox[boxsep=0.5mm, boxrule=0.5mm,  colframe=black, colback=white] 
            {\includegraphics[scale=0.4]{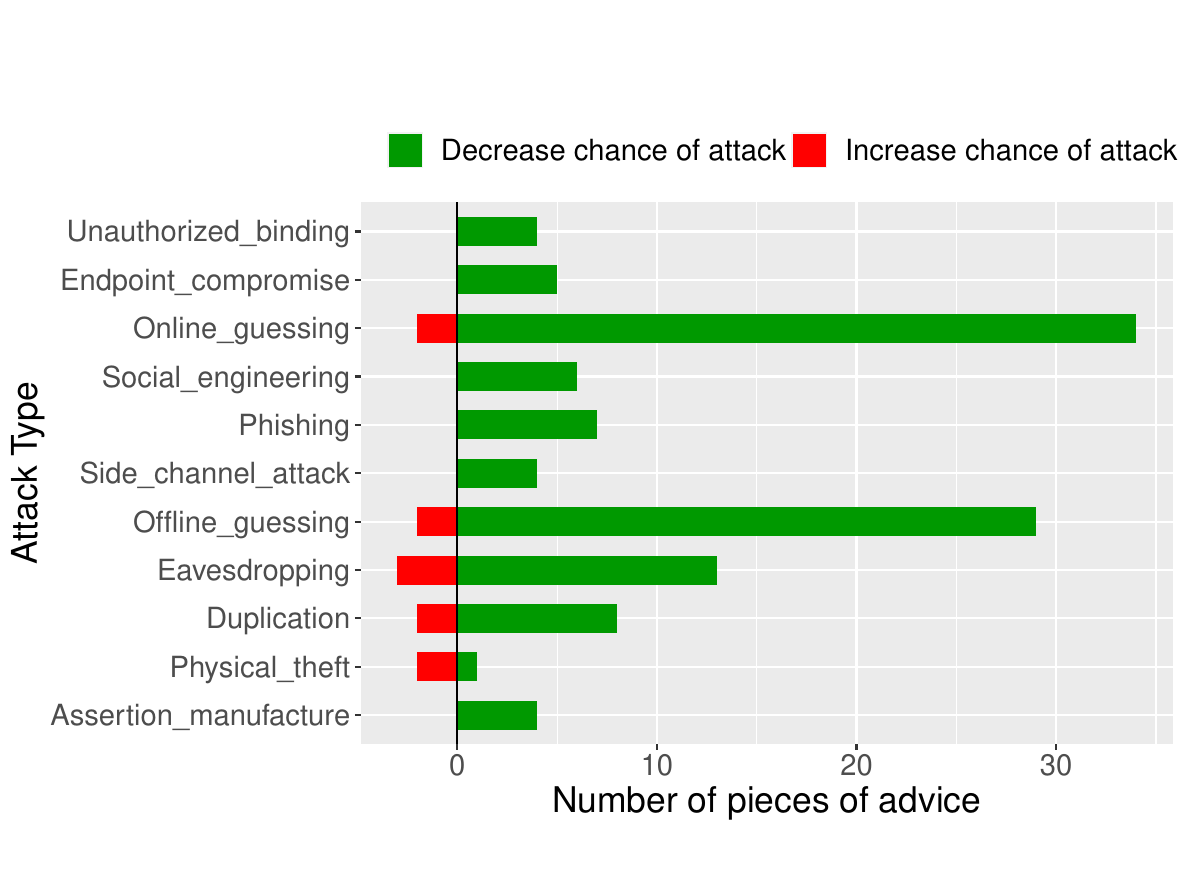}}
            \caption{\label{ben_stat}Number of times advice affects each attack type (excluding minor increases and decreases)} 
\end{figure}

Fig.~\ref{ben_stat} shows the number of times an attack was affected by a piece of advice. Green bars show the number of times a piece of advice decreased the chance of this attack occurring. The red-boxed bars show the number of times a piece of advice increased the chance of the given attack occurring.  This figure shows that the advice does not equally consider all the attack types. Most of the advice is focused on online guessing and offline guessing. This is a  reflection of the quantity of advice that focused on how a password should be created. Eavesdropping is the third most commonly protected against attack type. Eavesdropping refers to both online eavesdroppers which encrypted communications protect against and shoulder surfing eavesdroppers who attempt to view a password as it is typed.

One interesting insight is that the amount of advice given does not necessarily seem to correspond to the severity or regularity of the attack type. Only 7 pieces of advice protect against phishing or pharming attacks. This is in comparison to the 34 pieces of advice that protect against online guessing. In fact, 6 pieces of advice protect against social engineering and 4 against side channel attacks. Both of which appear less common attacks than phishing.

Physical theft had the least amount of advice that helped protect against it and was also the attack that advice most commonly lead to increased exposure. This is because of advice such as ``use 2-factor authentication using a phone'' introduces an additional physical object into the authentication procedure and therefore opens up new potential for a physical theft attack.



%

%


\section{Costs versus benefits}\label{sec:costsvsbenefits} 

As we have now reviewed the costs and benefits independently, in this section we will compare costs versus benefits for each piece of advice that we collected. We use an elementary scoring system which assigns positive scores for each benefit a piece of advice provides and subtracts points for each cost it incurs. Naturally, this method does not give a true insight into all nuances of the costs and benefits and we discuss this throughout this section.

\subsection{Scoring Costs} \label{sec:costlyadvice}
\begin{figure}
      \centering
\begin{subfigure}[b]{0.48\textwidth}
      \centering
            \includegraphics[width=\textwidth, height=5cm]{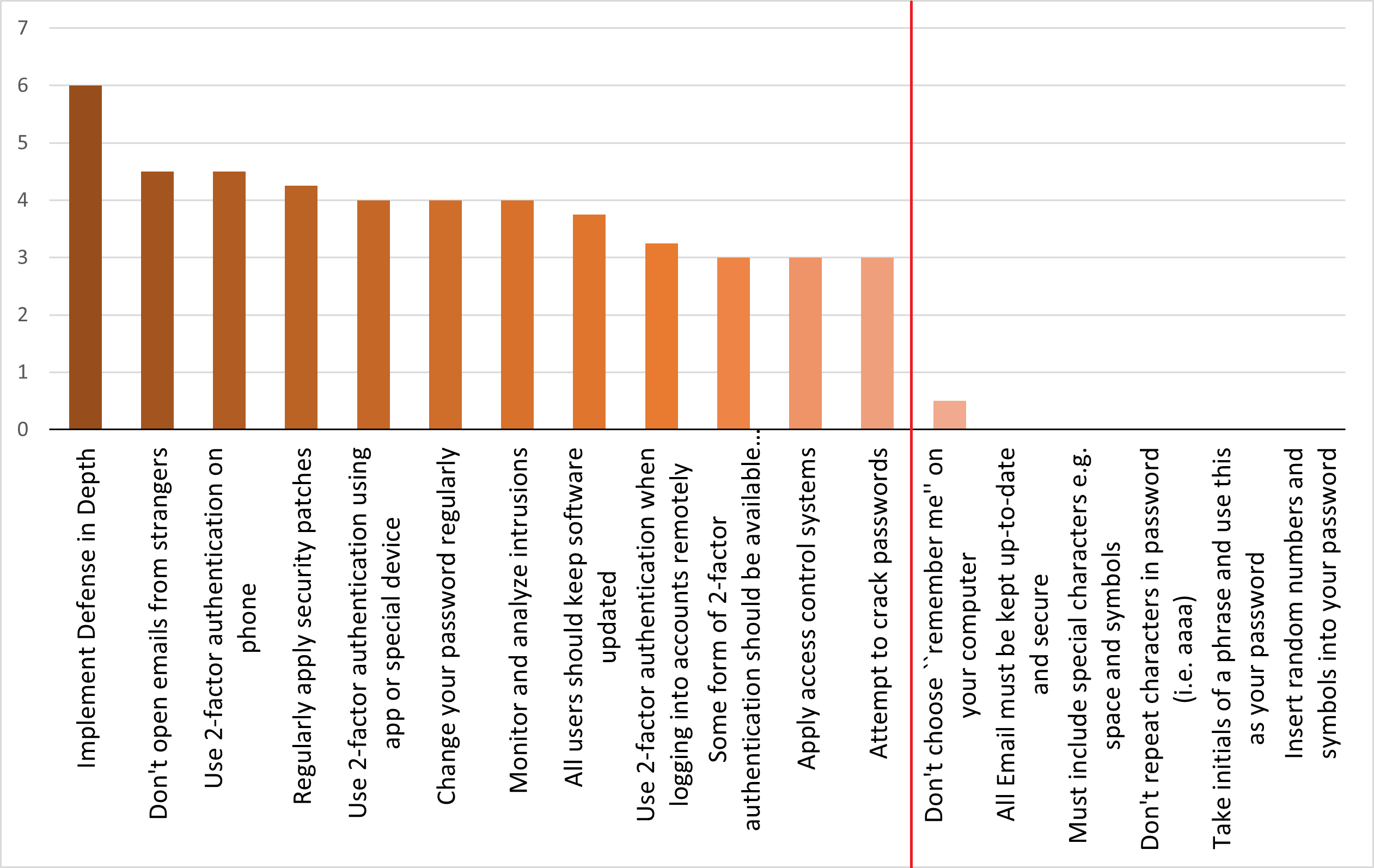}
      \caption{Costly advice for the organisation.}
      \label{fig:costly-org}
\end{subfigure}%
\hspace{1em}
\begin{subfigure}[b]{0.48\textwidth}\hspace{2em}
      \centering
            \includegraphics[width=\textwidth, height=5cm]{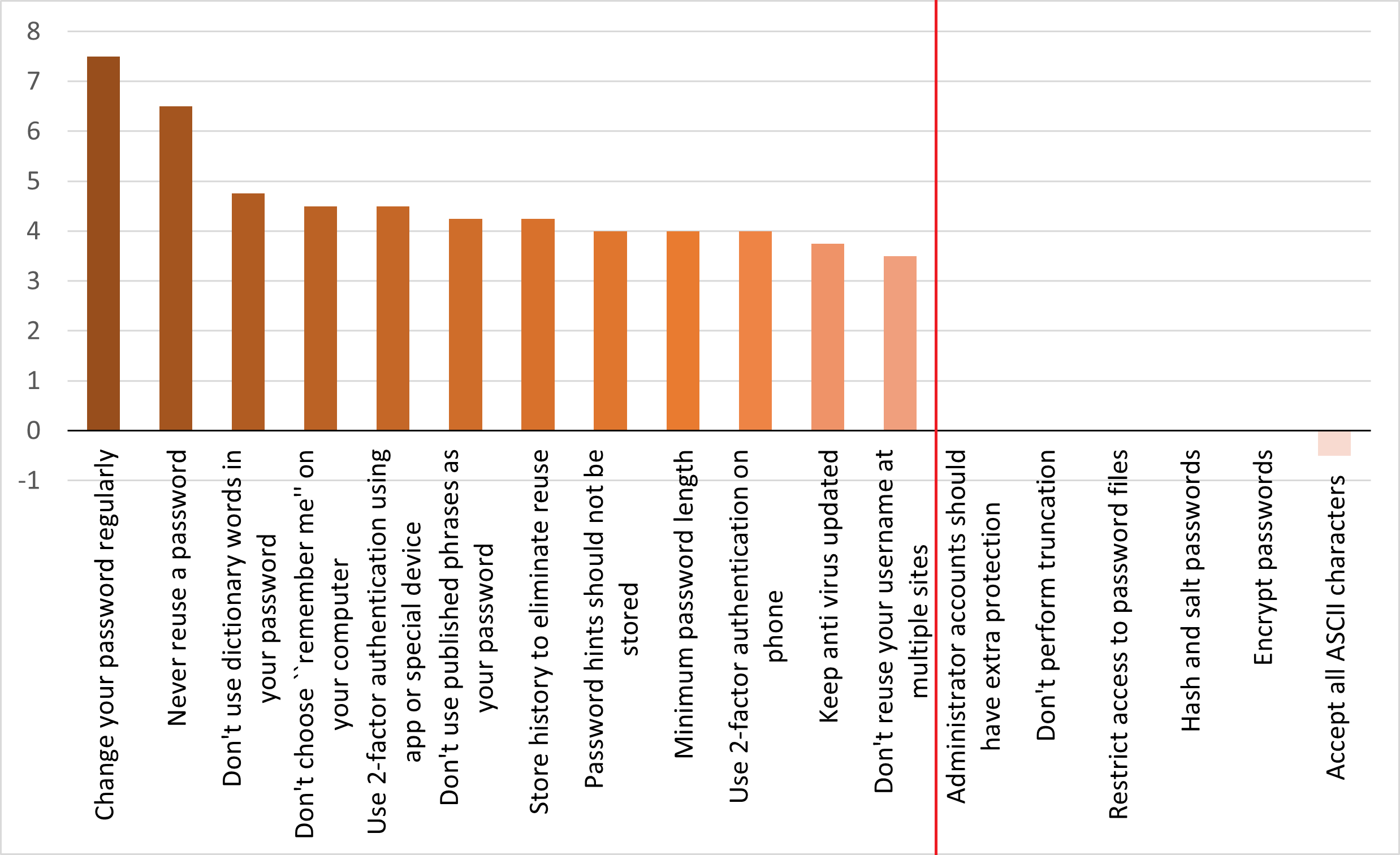}
      \caption{Costly advice for the end-users.}
      \label{fig:costly-user}
\end{subfigure}%
\caption{Bar chart showing the 12 most costly and 6 least costly pieces of advice for the organisation and for the user.}
\label{fig:advice-costs}
  \end{figure}

We begin by assigning points to each cost type. We assign 2 points to a major cost which reoccurs at each login, 1.5 to a major periodic cost, and 1 to a major one-time cost. A minor login cost is assigned 1, minor periodic cost 0.75, and a minor one time cost is 0.5. A positive cost can account for -2. This allows us to identify the most and least costly advice. Fig.~\ref{fig:advice-costs} depicts the 12 most costly pieces of advice and the 6 least costly pieces of advice for the organisation. 

In Fig.~\ref{fig:costly-org}, most of the least costly pieces of advice relate to composition of passwords. For example, ``Take initials of a phrase and use this as your password''. Giving advice list this does not bring any costs to the organisation. Many of the most costly pieces of advice for the organisation involve back end processes. For example ``Implement Defense in Depth'', ``Monitor and Analyse intrusions'' and ``apply access controls''.

Introducing two factor authentication shows up in both the user and the administrators most costly advice. Similarly the advice to change passwords regularly occurs in both top 12 lists. Where as, the advice ``Don't choose remember me on your computer'' is a high cost for users but is one of the low costs for the organisation. 

Most of the most difficult pieces of advice for users relate to password creation, memorizing and changing. For example ``Don't use dictionary words in your password'', ``Password hints should not be stored'', ``Store history to eliminate password reuse''. This last piece of advice specifically related to when regular password changes are enforced. It ensures a user does not choose a previous password. 

Notice that the last piece of advice in the user chart Fig.~\ref{fig:costly-user} has a negative cost. This is because, allowing users to use any ASCII characters actually makes their life easier. Of course, most security policies and rules will come at a cost. What will be interesting is to see whether the benefits outweigh these costs. 

\subsection{Scoring benefits}\label{sec:beneficialadvice}
For benefits we use the following rules: a large decrease in attack risk (\verydown) counts for 2 points, a minor decrease to attack risk (\down) is 1 point. A small increase to risk (\up) is -1 point and a large increase of risk (\veryup) is -2 points. 

Note that a comparison of benefits advice alone can only provide limited meaning. Some attacks are more important to protect against than others, and advice can protect against attacks to varying degrees. However, for the purpose of this analysis, a simple comparison of how many attacks a piece of advice protects against and whether it offers minor or major protection could be interesting. We therefore attempt a loose ranking of the advice according to the assigned benefits. `Security beneficial' is a simple indication of whether the advice protects against the different attacks from Tab.~\ref{tab:attacks}.

\begin{figure}
    \centering
    \includegraphics[width=0.7\linewidth]{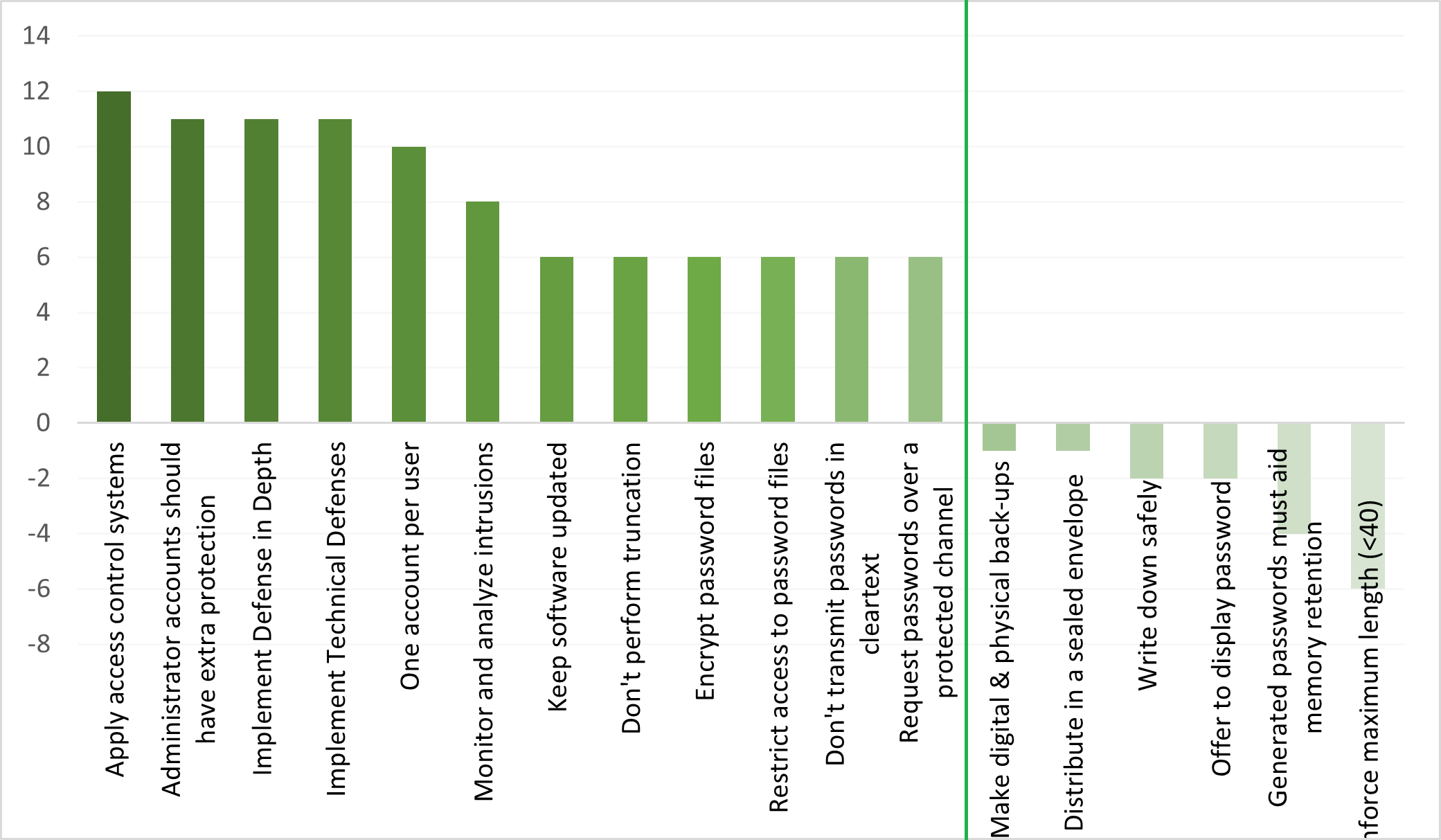}
    \caption{Bar chart showing the 12 most security beneficial and 6 least security beneficial pieces of advice. `Security beneficial' is a simple indication of whether the advice protects against the different attacks listed in Tab.~\ref{tab:attacks}.}
    \label{fig:advice-bens}
\end{figure}

Using the above scoring methodology, Fig.~\ref{fig:advice-bens} shows the 12 most security beneficial pieces of advice and the 6 least security beneficial pieces of advice. The advice that our simple quantification identifies as the most beneficial seems to be in line with what we might imagine. In fact, there is overlap between the 12 most beneficial pieces of advice and the 12 pieces of advice that had high organisation costs. For example, ``Apply access controls'', ``Implement defense in depth'' and ``monitor and analyse intrusions appear in both lists''. 

The six least beneficial pieces of advice actually result in negative benefits. This means they increase the likelihood of an attack occurring. Looking through the six pieces of advice we can understand why. For three of these advice statements, rather than increasing security, their goal seems to be to create a better user experience. For example, ``write [your password] down safely'', ``offer to display password'' and ``generated passwords must aid memory retention''. The other three pieces of advice have negative benefits because they create a new attack vector or make an attack more likely. Distributing passwords by envelope means they can be physically intercepted. Enforcing a maximum length puts an upper bound on the length of passwords and makes guessing attacks easier. Finally, though storing backups is an important secure practice, it rarely actually protects against attacks, instead it mitigates the harm done if an attack takes place. Therefore, purely in terms of protection from attacks, this ranks poorly. In addition, having both physical and digital copies creates extra data that must now be protected and physical protection is now a factor. This is a good example of how our simple points system doesn't give the whole picture of if a security practice should be employed.

\subsection{Costs versus benefits trade-off} 

We are interested in the trade-offs between the costs and the benefits. Is high-cost advice balanced by high benefits? Or are users paying high usability costs for small increases to security?  

\begin{figure}
    \centering
    \includegraphics[width=\linewidth]{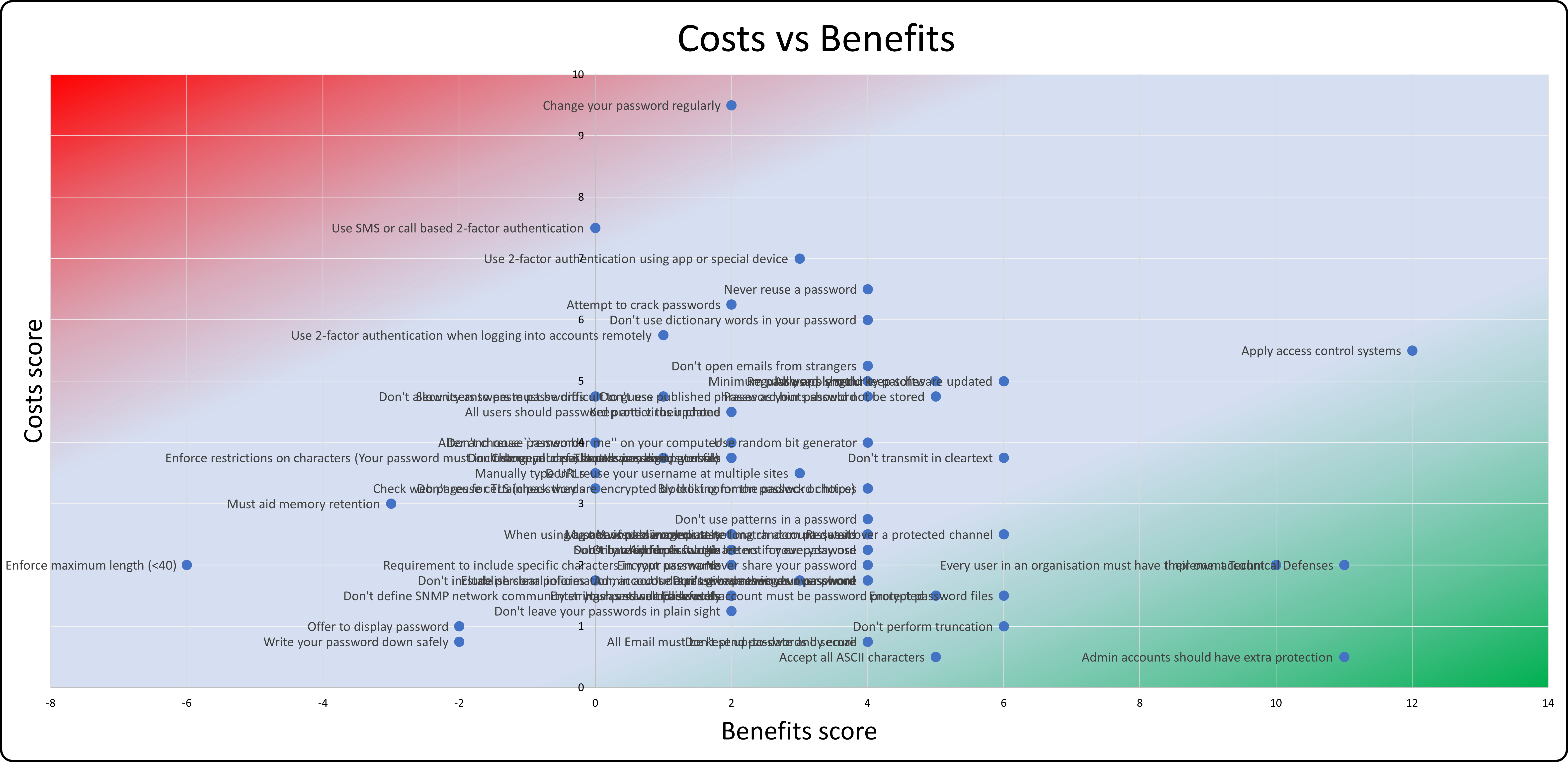}
    \caption{Scatter-plot comparing costs versus benefits of password advice. The red quadrant shows high cost low benefit advice. The green quadrant shows low cost high benefit advice. This graph is intended to give an impression of the distribution of the costs vs benefits and therefore it is not necessary to be able to read each point.}
    \label{fig:scatter-compare}
    \vspace{-2em}
\end{figure}

In Fig.~\ref{fig:scatter-compare} we plot a scatter-plot of every piece of advice. The benefit score is shown on the x-axis and the cost score on the y-axis. Advice will falls in the bottom right quadrant (green area) is high benefit and low cost advice. Similarly the advice that falls in the upper left quadrant (red area) in low benefit and high cost advice. The worst advice will have low benefits and high cost. Because we are particularly interested in the extremes advice, it does not matter that we cannot see the text of the centre section. However, if the reader is interested, we have provided the data and the graph in GitHub for a more detailed perusal~\cite{hazel_github}.

\subsubsection{High cost - low benefit advice}
Let us look first at the high cost low benefit advice. Reassuringly there is not too much advice in this quadrant. The only piece of advice that falls in the red area is ``change your passwords resularly''. This was high cost advice for both the end-user and the organisation. Research has also shown that it has few security benefits~\cite{chiasson2015quantifying,zhang2010security}. The NIST advice advice explicitly states that regular password change should not be enforced as it does more harm than good~\cite{nist2017}.

The next piece of advice closest to the top left (high cost - low benefit) quadrant is the advice ``Use SMS or call based 2-factor authentication''. While we have generally come to accept that two factor authentication offers security benefits, SMS text and phone based authentication has security flaws. In particular, text and phone calls are not protected by encryption and phone numbers can be easily spoofed. In the initial version of the NIST Digital Guidelines, SMS based two-factor authentication was to be deprecated~\cite{nist2016draft}. However, this decision was overturned~\cite{nist2017}. This newly introduced attack vectors explain why this advice receives a lower security ranking that 2FA using an app or specialised device in our plot. However, the general low ranking of 2FA in general is a reflection of the limitations of the model. Despite offering major decreased in attack probability in the important areas of phishing and online guessing, 2FA using a phone introduces four new attack vectors: Physical theft, eavesdropping, side channel attack and endpoint compromise. Since our model doesn't have probabilities or more graded weightings, it seems to think that these new attack vector outweigh the benefits. This is only an issue for the 10 pieces of advice that have both positive and negative security impacts, however, it is important to keep in mind.

\subsubsection{Low benefit advice}
In the bottom left quadrant, we can see low cost and low benefit advice. Enforcing a maximum password length falls directly into this category. ``Enforce maximum [password] length (<40)'' had no positive security value. This is an example of advice which compromises usability for no increase in security. Unfortunately, this practice is still enforced by organisations. In 2014, Saini assessed the policies of 23 different websites and 8 enforced maximum lengths of 40 characters or less on passwords~\cite{saini2014analysis}. Three of these limited the length to just 16 characters and one limited it at 12 characters. In the course of our study, we found that some websites only reveal their limit on password length after the user has attempted to use a longer password \cite{paypal}.

\subsubsection{High benefit advice}
``Administrator accounts should have extra protection'', ``Every user in an organisation must have their own account'' and ``Implement technical defenses'' all fall solidly within the bottom right quadrant. This is the advice that will bring high benefits for low costs. Other advice in this quadrant is ``Accept all ASCII characters'', ``Don't perform truncation'', ``Encrypt password files'', ``Each user account must be password protected'', ``Don't send passwords by email'', and ``All email must be kept up-to-date and secure''. ``Apply access control systems'' comes at a higher cost but offers the strongest benefits of any advice.





\subsubsection{Some benefit - some cost}
The advice in the blue diagonal, is more difficult for us to form conclusions about. Anything on the positive x-axis offers security benefits, but inevitably comes at a cost. For this advice, a more detailed quantitative analysis would be necessary in order to determine whether benefits outweigh costs. 

For example, one expensive piece of advice is ``Use 2-factor authentication using app or special device''. However, it offers increases in security against online guessing and phishing attacks, two of the most common attack types. It is likely that ensuring the benefits outweigh the costs will depend on its implementation and the needs of the specific organisation and users. These nuances are something our current model can't uncover. 

Similarly, the piece of advice ``Use a Password manager'' lies at \verb|(benefits=4,costs=2)| on the plot. It was well regarded by users in our study. It can protect against three attack types and most of the costs it incurs are to the organisation. A password manager greatly reduces the users' memory load and by extension a user can use as long, random and complex of a password as they wish. However, as with many of the pieces of advice, the value of a password manager will lie in how users utilise it. If a user uses a password manager and continues to reuse a common password choice across multiple sites then many of the potential benefits won't materialize.

As expected our ``canary'' piece of advice ``Do not allow users to paste passwords'' does not have a security benefit. However it did come with costs (at \verb|(benefits=0,costs=4.75)| in Fig.~\ref{fig:scatter-compare}). We found it surprising that it was not an outlier in the scatterplot. Instead, it is within a cluster of advice at the center of the scatter plot representing advice with costs but limited benefits. This, perhaps, tells us something about security advice that is not carefully sourced.

\subsection{Discussion}
We see that, in many cases cases, it is difficult to distinguish whether the costs outweigh the benefits for security advice at a glance. Most advice had some negative impact for users of the organization. This difficulty in assessing the trade-off could evidence one of the reasons why users and organizations often follow advice which researchers have proven ineffective. It seems, an `at a glance' observation, even by a security professional, might not always be possible for many of the pieces of advice we collected. Interestingly, we would have expected the high benefit advice to correlate with high costs. However, when we plotted a trend line for the Scatterplot in Fig.~\ref{fig:scatter-compare} this was not the case. The trend line in fact had a negative slope of -0.0373. The fact that the high cost advice does not map to the high benefits, highlights that the advice we force users and organisations to follow does not necessarily result in positive returns for their effort.



%


\section{Summary of results}\label{sec:summ}

Using our taxonomy of 270 pieces of collected and categorised password advice, we develop a model to determine the costs associated with enforcing and following security advice. We identify the costs of authentication advice as the resources, human or otherwise, which are required for the advice to be implemented. This method involved assigning a severity (major, minor, positive) and frequency (once-off, periodic, at login) to each cost category for each piece of advice. Costs could be positive or negative, e.g. advice can reduce the number of resources or time needed.  After developing this methodology we asked 73 end-users and administrators for their opinions on the categories of costs we had identified and to indicate which costs they associated with each piece of security advice. While users and administrators largely agreed that the model we created is an accurate way of differentiating costs, it made for a convoluted user study. Some participants misunderstood the ranking, in particular the concept of a `positive cost' being one that has a positive impact on their authentication process. We added in clarification part way through and removed 3 answers where users clearly user `positive cost' to indicate that they were positive there was a cost. If repeating this survey, we would simplify the general questionnaire and potentially conduct detailed interviews with a smaller cohort of participants where we can learn about the finer details. On the whole, we still received meaningful feedback from our participants.

The survey highlighted that most of the security advice collected places a large burden on humans, both system administrators and end-users. Over 85\% of the costs we identified related to the need for additional human labour or effort. Human were impacted by 87\% of all the costs identified and 44\% of these were repeating costs, for example, at every login it adds on extra time. In the user survey we witnessed strong emotional responses towards the advice we asked users to analyse. We received responses such as ``I am sick of passwords and logins and they are making me less productive as I have to look up the passwords so often!!''. 

As part of the user survey, we also asked both groups whether they approved of each advice statement or not. Strikingly, we found end-users disagreed with each other 71\% of the time about whether a piece of advice was valuable or not. Along with the contradiction in the advice that is circulated by organisations, this shows serious cohesion issues within user-education and professional's security understanding. 

Once we had identified the costs of the security advice, we sought to qualify what benefits this advice offered. We defined the benefits as the change in security risk. Like costs, benefits can also be positive or negative i.e. advice can increase or decrease the risk of an attack. We used the NIST 2017 Digital Identity Guidelines list of authenticator threats as our set of possible attacks. For each attack, we indicated whether each piece of advice increased or decreased the likelihood of the attack occurring. We validated this assignment of security benefits by gathering input from various specialists within our university and at specialist conferences (\cite{usenix17_posters, murray2017evaluating, Passwordscon_talk,HEAnet_talk})\footnote{In supplementary material we include all our explanations for why particular security benefits were aligned to each piece of advice~\cite{hazel_github}.}. Despite this, this is inherently a subjective assessment based on our own knowledge and available research. More sophisticated threat and vulnerability assessment of risk reduction could have provided more concise assessments of security benefit. However, such a detailed qualification comes with its own problems and is beyond the scope of our simple assessment structure. 

From our assessment of the advice benefits, we found that most of the advice was concerned with protecting against online and offline guessing attacks. Keylogging and phishing or pharming attacks are not mentioned as often, despite their prevalence. Also, most of the protection against offline guessing was focused on improving password strength rather than back end-processes. This is an example of organisations placing the security burden on the user, rather than  putting in proper protections themselves.

In our final section, we compared the costs to the benefits for each piece of security advice. We found that the advice that was most costly to users was different to the advice that was most costly to the organisation. We also found that the most security beneficial advice overlapped with the advice that had high organisation costs. This shows that if organisations are willing to make an investment in security, then they can implement advice which will result in strong security benefits. The same is not true for user advice. 

While we could identify a selection of advice that seemed valuable (high benefit and low cost) and one piece of advice that was not worthwhile (low benefit and low cost), for the majority of the advice it was unclear whether benefits did outweigh costs. We conclude that quick assessments of whether advice is valuable are not always possible. In fact, this is emphasised through out the paper. We found contradictions in the advice websites and organisations circulate and enforce. We found that end-users found it difficult agree on what advice they approved of. Clearly it is a difficult challenge for organisations and users to determine the nuances of whether security benefits outweigh the costs of advice. This difficulty in assessing whether security advice is valuable could help explain the incongruity of advice and opinions in the wild.


%

 %
 %

\section{Conclusion}\label{sec:conc} 
In this paper, we collected 270 pieces of authentication security advice that were given by security specialists, multinational companies and public bodies. This collection highlighted stark variations between advice given by different sources. 41\% of the recommendations we collected were contradicted by recommendations given by another source. We surveyed administrators and users about the costs associated with following this advice. The results of this study are tabulated and illustrate the costs and perceptions that accompany password advice. Our research exposed the disconnect that exists between the recommendation of security research, and the advice given by organisations and believed by users. Finally, we qualify which attack vectors each piece of advice offered protection against. We find that most advice was concerned with protecting against online and offline guessing attacks, with little emphasis on the security of back-end processes. We also find that some advice that is circulated has no discernible security value and also hinders usability. This research highlights the need for organisations to follow best practices guidelines when giving advice rather than preconceived notions of what should be secure.

\begin{acks}
 H Murray was supported by an Irish Research Council 2017 Government of Ireland Postgraduate Scholarship and a John and Pat Hume doctoral studentship.
 This publication was supported in part by a research grant from Science Foundation Ireland (SFI) and is co-funded under the European Regional Development Fund under Grant 13/RC/2077\_P2.
\end{acks}

\FloatBarrier

\bibliographystyle{ACM-Reference-Format}
\bibliography{bib5}


\end{document}